\documentclass[twocolumn]{aastex62}

\usepackage{graphicx}
\usepackage{placeins}
\usepackage{amsmath}

\received{June 6, 2019}
\revised{August 6, 2019}
\accepted{August 20, 2019}
\submitjournal{ApJ}

\shorttitle{HPS [O~III] survey}
\shortauthors{Indahl et al.}

\begin{document}

\title{HETDEX Pilot Survey. VI. [O~III] Emitters and \\
   Expectations for a Local Sample of Star Forming Galaxies in HETDEX}

\correspondingauthor{Briana Indahl}
\email{blindahl@astro.as.utexas.edu}

\author{Briana Indahl}
\affiliation{Astronomy Department, University of Texas at Austin, Austin, TX 78712}
\nocollaboration

\author{Greg Zeimann}
\affiliation{Hobby Eberly Telescope, University of Texas, Austin, Austin, TX, 78712}
\nocollaboration

\author{Gary J. Hill}
\affiliation{Astronomy Department, University of Texas at Austin, Austin, TX 78712}
\affiliation{McDonald Observatory, University of Texas, Austin, TX, 78712}
\nocollaboration

\author{Steven L. Finkelstein}
\affiliation{Astronomy Department, University of Texas at Austin, Austin, TX 78712}
\nocollaboration

\author{Robin Ciardullo}
\affiliation{Department of Astronomy \& Astrophysics, Pennsylvania State University, University Park, PA 16802}
\affiliation{Institute for Gravitation \& the Cosmos, Pennsylvania State University, University Park, PA, 16802}
\nocollaboration

\author{Joanna S. Bridge}
\affiliation{Department of Physics and Astronomy, 102 Natural Science Building, University of Louisville, Louisville, KY 40292}
\nocollaboration

\author{Taylor Chonis}
\affiliation{Ball Aerospace, Boulder, CO 80301}
\nocollaboration

\author{Niv Drory}
\affiliation{McDonald Observatory, University of Texas, Austin, TX, 78712}
\nocollaboration

\author{Caryl Gronwall}
\affiliation{Department of Astronomy \& Astrophysics, Pennsylvania State University, University Park, PA 16802}
\affiliation{Institute for Gravitation \& the Cosmos, Pennsylvania State University, University Park, PA, 16802}
\nocollaboration

\author{Hanshin Lee}
\affiliation{McDonald Observatory, University of Texas, Austin, TX, 78712}
\nocollaboration

\author{Kristen McQuinn}
\affiliation{McDonald Observatory, University of Texas, Austin, TX, 78712}
\nocollaboration

\begin{abstract}
We assemble an unbiased sample of 29 galaxies with [O~II] $\lambda 3727$ and/or [O~III] $\lambda 5007$ detections at $z < 0.15$ from the Hobby-Eberly Telescope Dark Energy Experiment (HETDEX) Pilot Survey (HPS\null).  HPS finds galaxies without pre-selection based on their detected emission lines via integral field spectroscopy. Sixteen of these objects were followed up with the second-generation, low resolution spectrograph (LRS2) on the upgraded Hobby-Eberly Telescope. Oxygen abundances were then derived via strong emission lines using a Bayesian approach. We find most of the galaxies fall along the mass-metallicity relation derived from photometrically selected star forming galaxies in the Sloan Digital Sky Survey (SDSS)\null. However, two of these galaxies have low metallicity (similar to the very rare green pea galaxies in mass-metallicity space). The star formation rates of this sample fall in an intermediate space between the SDSS star forming main sequence and the extreme green pea galaxies. We conclude that spectroscopic selection fills part of the mass-metallicity-SFR phase space that is missed in photometric surveys with pre-selection like SDSS, i.e., we find galaxies that are actively forming stars but are faint in continuum. We use the results of this pilot investigation to make predictions for the upcoming unbiased, large spectroscopic sample of local line emitters from HETDEX\null. With the larger HETDEX survey we will determine if galaxies selected spectroscopically without continuum brightness pre-selection have metallicities that fall on a continuum that bridges typical star forming and rarer, more extreme systems like green peas.


\end{abstract}




\section{Introduction}
\label{sec:intro}

The chemical properties of a galaxy constrain the history of its evolution. Metals in galaxies are a direct product of star formation. As stars form in a galaxy, supernovae spread metals that mix with the interstellar medium, and in turn, these metals allow for cooling of gas and more star formation. However, intense starbursts also drive gas containing metals out of a galaxy, which, depending on a galaxy's mass, may or may not be reaccreted. Primordial gas will also stream into galaxies diluting their metals. The metallicity of a galaxy is an observable  quantity that directly probes this recycling of baryonic material \citep[e.g.,][]{mai19}.
    
The idea that the metal content of galaxies is correlated with their mass and luminosity started with early studies such as \citet{van68} and \citet{pei70} using small samples of local group galaxies. Later studies, such as \citet{fab73} and \cite{leq79}, found similar correlations with mass and metallicity in elliptical and irregular galaxies, respectively. Subsequent investigations produced similar findings using small galaxy samples, but it was the release of the Sloan Digital Sky Survey (SDSS) which solidified the mass-metallicity relation (MZR) and defined a tight ($\sim 0.1$~dex) correlation for nearly $\sim 100,000$ local star forming galaxies \citep{tre04}. Many studies followed, which refined the local MZR for typical star forming galaxies, improved the techniques for measuring metallicity, and extended the relation to include the effects of star formation rate (SFR\null).  This new 3-parameter relation has been called the fundamental metallicity relation (FMR) \citep{man10, per13, and13}. The MZR has been derived for many unique nearby galaxy populations, such as ultra low mass galaxies \citep{lee06, ber12}, Lyman-break galaxy analogs \citep{lia15}, ``green pea" galaxies \citep{car09, izo11, haw12}, and ``blueberry" galaxies \citep{yan17}. Green pea and blueberry galaxies are selected in broad-band imaging surveys to have large [O~III] $\lambda5007$ equivalent widths. 

All of these previous studies of the mass, metallicity, and SFR of local galaxies derive their samples from continuum photometric selection. Metallicities for the star forming galaxies in SDSS are measured from spectra but the objects themselves were originally selected using photometry of the red continuum from older stars, as large area spectroscopy is typically very expensive. 
But these samples are biased against star forming galaxies that are faint in continuum. It is unknown how or if this bias might be manifested in widely accepted relationships between galaxy properties, such as the MZR and star forming main sequence. 

Spectroscopic surveys without target pre-selection were originally made possible with slitless objective prism and grism surveys such as the Universidad Complutense de Madrid (UCM) survey \citep{gal96} and the KPNO International Spectroscopic Survey (KISS) objective-prism survey \citep{sal00} on the ground. In space the Advanced Camera for Surveys (ACS) conducted grism surveys such as the Grism ACS Program for Extragalactic Science (GRAPES) \citep{pir04} and the 3D-HST grism survey \citep{van11, bra12}. However, extraction of emission line objects, especially in crowded fields, can be very challenging and these surveys are often background-limited.  Moreover, while KISS is an objective-prism emission line survey, it relied on continuum selection to extract their objects, meaning even strong emission-line objects would be missed if their continuua were too faint \citep{sal00}. More advanced techniques have been developed for ACS grism data \citep{meu07, mas18a}; however, these surveys mostly focus on high redshift galaxy samples and have a very small survey area. 

Integral field spectroscopy has made it easier to conduct deeper and more complete spectroscopic surveys of galaxies, as the technique greatly reduces and the background and avoids the inconveniences associated with overlapping spectra. For the most part, however, these instruments have a small field of view, which limits their effectiveness to either small areas, such as \citep[i.e., the Multi Unit Spectroscopic Explorerv (MUSE)][]{mas18b}, or to rely on pre-selection of targets, such as the Calar Alto Legacy Integral Field Area Survey (CALIFA) \citep{san12},  Mapping Nearby Galaxies at APO (MaNGA) \citep{bun15}, the Sydney-Australian-Astronomical-Observatory Multi-object Integral-Field Spectrograph (SAMI) \citep{bry15}, and Hector \citep{bry16} programs.

The wide field upgrade to the Hobby Eberly Telescope at McDonald Observatory \citep[HET;][]{hil18b} and the $\sim 35,000$ $1\farcs 5$-diameter fibers of the Visible Integral Field Replicable Unit Spectrograph's \citep[VIRUS;][]{hil18a} changes this. Specifically, the Hobby-Eberly Telescope Dark Energy Experiment (HETDEX) utilizes VIRUS to provide an extremely large sample of galaxies selected purely on their line emission.   This survey will blindly sample 420~deg$^2$, divided by a 4.5 fill factor to account for spacing between fiber bundles, over a 5 year period.   The motivation of the HETDEX survey is to sample Ly$\alpha$ galaxies between $1.9< z < 3.5$ in order to measure the expansion rate of the universe at that epoch. However, due to its blind nature, the experiment will detect star-forming galaxies at a variety of redshifts, via emission-lines as faint as $3.5 \times 10^{-17}$~ergs~cm$^{-2}$~s$^{-1}$ monochromatic flux limit \citep{hil08}.\par

In preparation for HETDEX, a pilot study was conducted with the prototype instrument, the George and Cynthia Mitchell Spectrograph (GMS; previously known as VIRUS-P) on the McDonald Observatory 2.7 meter Harlan J. Smith Telescope \citep{ada11, bla11}. The HETDEX Pilot Survey (HPS) was also a blind spectroscopic survey covering an area of $\sim 163$~arcmin$^2$. To understand the population of galaxies that HETDEX will detect, we have conducted a pilot study of a sample of local star forming galaxies selected from the HPS data, and present initial results and expectations for the larger HETDEX sample. Specifically, we assemble a sample of 29 galaxies with $z < 0.15$ found either through detection of the [O~II] $\lambda 3727$ or the [O~III] $\lambda 5007$ forbidden lines. Using these strong lines and H$\beta$, we implement a method of measuring metallicity which utilizes a Bayesian approach to fitting strong-line relations. \citet{gra16} developed this approach to measure the metallicities of a sample of intermediate redshift ($1.90 < z < 2.35$) galaxies where constraints from redder lines such as [N~II] $\lambda 6583$ and H$\alpha$ are unavailable. We use this same technique for HPS galaxies (and by extension HETDEX), where the spectral range of the instrument does not allow for measurements of the longer wavelength lines.  With this sample we explore the phase space in mass, metallicity, and SFR that is potentially missing in surveys selected by continuum photometry. This paper presents a pilot study of the methods and expectations for the larger sample of [O~III] emitting galaxies that will be found in the HETDEX database.\par

Section \S~\ref{sec:data} 
of this paper summarizes the HPS sample by first outlining the survey and describing how our galaxies were identified (\S~\ref{sec:hps}). This section also details our follow up observations with the second generation low resolution spectrograph (LRS2) aimed at obtaining deeper spectroscopy on a subset of the sample (\S~\ref{sec:lrs2_obs}), as well as the reduction (\S~\ref{sec:lrs2_redux}), and relative flux calibration of these data (\S~\ref{sec:rel_cal}).  In Section \S~\ref{sec:fluxes} we outline our procedure to measure our LRS2 line fluxes (\S~\ref{sec:elf}), correct for underlying H$\beta$ stellar absorption (\S~\ref{sec:hb_corr}), and obtain an absolute calibration (\S~\ref{sec:abs_cal}).  Section \S~\ref{sec:met} compares measures of metallicity and describes the metallicity fitting routine and Section \S~\ref{sec:mass} describes how we obtain stellar mass estimates for our galaxies.  In Section \S~\ref{sec:MZR} we describe how our spectroscopically selected sample compares to well-established photometrically-selected galaxy populations in terms of the mass-metallicity relation. SFR rates are discussed in Section \S~\ref{sec:sfr} as well as the SFR vs.\ mass relation. Implications of this study for the larger HETDEX sample are discussed in Section \S~\ref{sec:hetdex} and Section \S~\ref{sec:conclusions} summarizes our findings. \autoref{sec:ql} provides a description of the LRS2 Quick Look Pipeline, which is a code used for the reduction of LRS2 data. 

In this paper we adopt a $\Lambda$CDM cosmology with $H_0 = 70$~km~s$^{-1}$~Mpc$^{-1}$, $\Omega_M=0.3$, and $\Omega_{\Lambda}=0.7$. References to metallicity refer to the gas-phase  oxygen abundance, which is generally given by its ratio to hydrogen, i.e., 12+log(O/H). 

\section{Sample Selection and Data}
\label{sec:data}

The [O~II] $\lambda 3727$ and [O~III] $\lambda 5007$ emitting galaxies in our sample were selected from the blind, spectroscopic HETDEX Pilot Survey discussed in \S~\ref{sec:hps}.  Some objects with lower signal-to-noise emission lines required follow up with LRS2.  These are discussed in  \S~\ref{sec:lrs2_obs}.

\subsection{HETDEX Pilot Survey Sample}
\label{sec:hps}

\begin{deluxetable*}{lcccccc}
\centering
\tablecaption{HPS objects and their properties derived from the HPS catalog \cite{ada11} \label{objs_tbl}}
\tablehead{\colhead{HPS name} & \colhead{HPS ID} & \colhead{Selection} & \colhead{$z$} & \colhead{$\alpha(2000)$} & \colhead{$\delta(2000)$} & \colhead{Field}}
\startdata
HPS022127-043019 & 35 & [OII]3727 & 0.0841 & 35.364375 & -4.505306 & XMM-LSS \\
HPS030630+000128 & 44 & [OII]3727 & 0.1191 & 46.628042 & 0.024667 & MUNICS-S2 \\
HPS030638+000015 & 65 & [OII]3727 & 0.1121 & 46.661458 & -0.004417 & MUNICS-S2 \\
HPS030638+000240 & 67 & [OII]3727 & 0.1123 & 46.662125 & -0.044472 & MUNICS-S2 \\
HPS030649+000314 & 105 & [OII]3727 & 0.1094 & 46.707250 & -0.054000 & MUNICS-S2 \\
HPS030651-000234 & 118 & [OII]3727 & 0.1072 & 46.716417 & -0.042972 & MUNICS-S2 \\
HPS030652+000123 & 119 & [OII]3727 & 0.1121 & 46.716750 & 0.023167 & MUNICS-S2 \\
HPS030655-000050 & 125 & [OII]3727 & 0.1118 & 46.729583 & -0.013972 & MUNICS-S2 \\
HPS030655+000213 & 129 & [OII]3727 & 0.1377 & 46.732542 & 0.037083 & MUNICS-S2 \\
HPS030657+000139 & 138 & [OII]3727 & 0.1027 & 46.740125 & -0.027528 & MUNICS-S2 \\
HPS100008+021542 & 158 & [OII]3727 & 0.0929 & 150.035500 & 2.261833 & COSMOS \\
HPS100018+021818 & 219 & [OII]3727 & 0.1231 & 150.075417 & 2.305194 & COSMOS \\
HPS100018+021426 & 225 & [OII]3727 & 0.1243 & 150.078917 & 2.240722 & COSMOS \\
HPS100021+021351 & 234 & [OIII]5007 & 0.0917 & 150.089542 & 2.230972 & COSMOS \\
HPS100021+021223 & 235 & [OII]3727 & 0.1317 & 150.090292 & 2.206556 & COSMOS \\
HPS100021+021237 & 237 & [OII]3727 & 0.1230 & 150.091417 & 2.210389 & COSMOS \\
HPS100028+021858 & 260 & [OII]3727 & 0.1063 & 150.119042 & 2.316222 & COSMOS \\
HPS100032+021356 & 278 & [OII]3727 & 0.1226 & 150.136875 & 2.232444 & COSMOS \\
HPS100037+021254 & 300 & [OII]3727 & 0.1159 & 150.156333 & 2.215167 & COSMOS \\
HPS100039+021246 & 303 & [OII]3727 & 0.1448 & 150.163875 & 2.212917 & COSMOS \\
HPS100045+021823 & 326 & [OII]3727 & 0.1222 & 150.188417 & 2.306500 & COSMOS \\
HPS123632+621037 & 363 & [OII]3727 & 0.1362 & 189.135333 & 62.176917 & GOODS-N \\
HPS123636+621135 & 375 & [OII]3727 & 0.0786 & 189.152958 & 62.193139 & GOODS-N \\
HPS123641+621131 & 386 & [OII]3727 & 0.0891 & 189.173708 & 62.192139 & GOODS-N \\
HPS123648+621426 & 413 & [OII]3727 & 0.1389 & 189.200958 & 62.240750 & GOODS-N \\
HPS123652+621125 & 430 & [OIII]5007 & 0.0889 & 189.216792 & 62.190528 & GOODS-N \\
HPS123656+621420 & 438 & [OII]3727 & 0.1053 & 189.235417 & 62.238750 & GOODS-N \\
HPS123659+621404 & 449 & [OII]3727 & 0.0887 & 189.247917 & 62.234694 & GOODS-N \\
HPS123702+621123 & 458 & [OII]3727 & 0.1359 & 189.258917 & 62.189917 & GOODS-N
\enddata
\end{deluxetable*}

The 29 low redshift galaxies in our sample were selected from the HPS Emission Line Catalog \citep{ada11}. The George and Cynthia Mitchell Spectrograph \citep[GMS, aka VIRUS-P;][]{hil08} was constructed specifically for the 2.7 meter Harlan J. Smith Telescope as a prototype of a single channel of the VIRUS instrument that is currently being used for HETDEX\null.  GMS is a large field ($1\farcm 7 \times 1\farcm 7$) integral field unit (IFU) spectrograph fed by 246 200$\micron$-core, densepak fibers; at the focal plane of the telescope, each fiber core projects to $4\farcs 2$ arcsec diameter on sky.  For the HPS survey GMS was configured to cover $\sim 3500$ to 5800~\AA\  at $\sim 5$~\AA\ FWHM resolution.  The HPS surveyed a total area of 163.23 arcmin$^2$ using 27 pointings in the Cosmological Evolution Survey \citep[COSMOS;][]{sco07}, 16 pointings in the Munich Near-IR Cluster Survey \citep[MUNICS;][]{dro01}, 13 pointings in the Hubble Deep Field North \citep{wil96} and Great Observatories Origins Deep Survey North \citep[GOODS-N;][]{dic03}, and 4 fields in the XMM Large-Scale Structure field \citep[XMM-LSS;][]{pie04}.

\begin{figure*}[ht!]
\epsscale{1.0}
\plotone{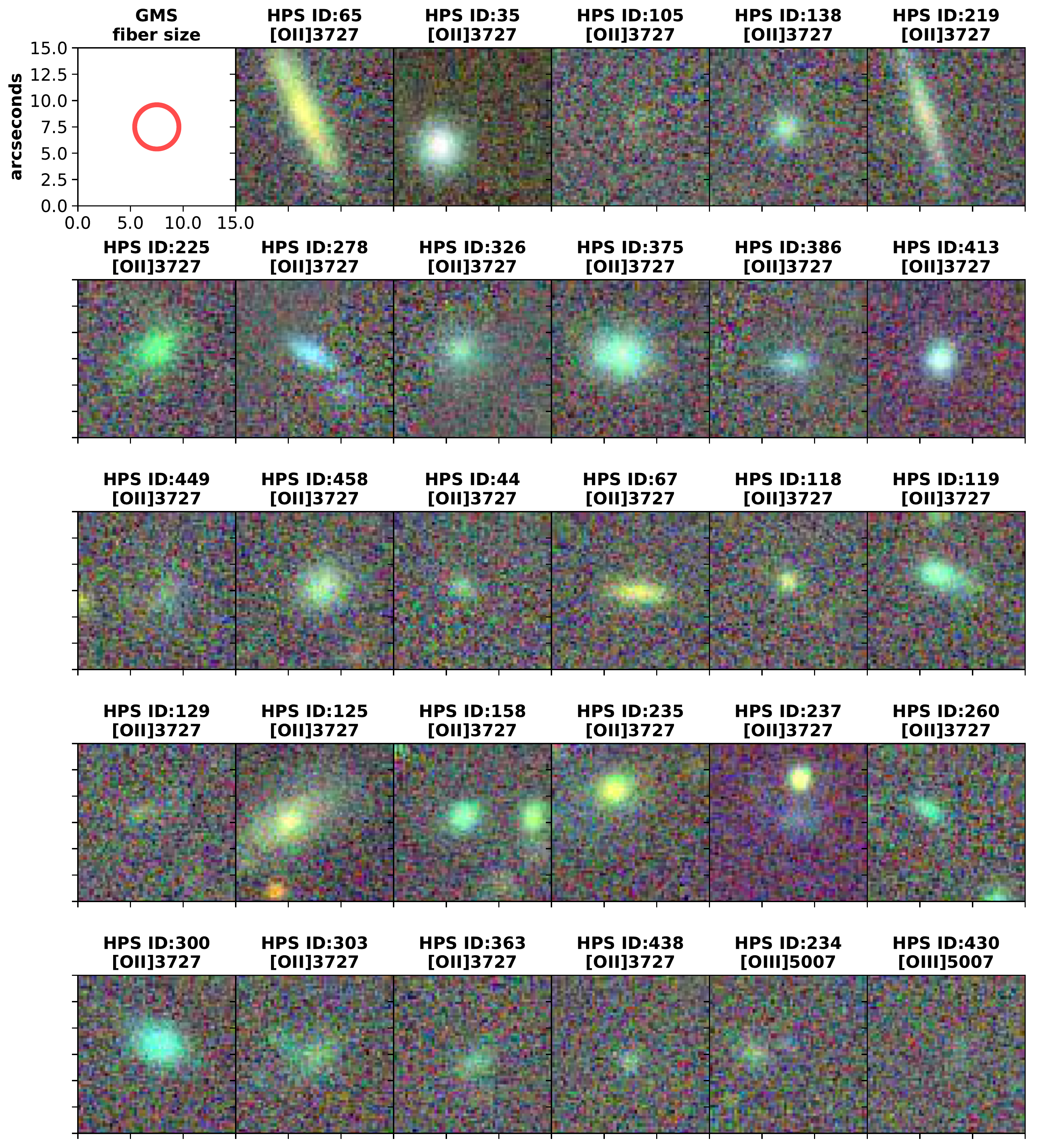}
\caption{PanSTARRS cutouts in $g$, $r$, and $z$ for the 29 galaxies in our sample.  Each cutout is $15\arcsec$ on a side, and is labeled with its object name, HPS catalog ID and the line by which it was selected. The first plot shows the size of the 4.2 arcsecond fibers in  GMS.}.\label{fig:cutouts}
\end{figure*}

The goal of HPS was to survey a patch of sky without target pre-selection in search of emission line galaxies, in order to test the principles of the HETDEX survey. As with VIRUS, GMS lacks the resolution to split the [O~II] doublet, so in order to distinguish between Ly$\alpha$ and lower redshift [O~II], HPS uses an equivalent width (EW) cut:  in the local universe, galaxies rarely have [O~II] rest-frame EWs above 20~\AA\ 
\citep{gal96, bla00, gro07, cia13}, while the vast majority of LAEs have rest-frame equivalent widths greater than 60~\AA\null. Since it was not expected that the continuum would be detected for most objects, ancillary imaging in the four fields was used to make the EW estimates. In total HPS found 397 unique emission line galaxies, including 285 [O~II] emitters with $z < 0.56$ and 105 Ly$\alpha$ emitting galaxies from $1.9 < z < 3.8$ \citep{ada11}.

In addition to the [O~II] $\lambda 3727$ line, strong-line metallicity estimates require the measurement of (at least) H$\beta$ and [O~III] $\lambda 5007$. To ensure all 3 lines are present, our sample only includes entries with $z < 0.15$ in the emission line catalog, as more distant systems will have [O~III] $\lambda 5007$ redshifted out of the spectral range of the survey.  We assembled a sample of all cataloged [O~II] $\lambda 3727$ and [O~III] $\lambda 5007$ emission line detections satisfying the redshift cut. 

The HPS emission line catalog contains a separate entry for every emission-line detection, so a single object may have multiple entries. To pair [O~II] $\lambda 3727$, [O~III] $\lambda 5007$ and H$\beta$ detections in the same object, we required that the lines agree in both redshift ($|\Delta z| < 0.001$) and the position ($|\Delta(\alpha,\delta)| < 3\arcsec$).
 
Photometric data from PanSTARRS \citep{cha16} were used to visually inspect each galaxy (see cutout images in \autoref{fig:cutouts}). Most objects appear to be fairly compact with a few being extended (the largest being about 15\arcsec).  Based on these inspections, a couple of line detections were confirmed to be from the same extended source, and were thus added to our sample. We also removed two probable AGN from our sample, as they had X-ray counterparts in \texttt{Chandra} and \texttt{XMM} survey data \citep{ada11}.

\begin{figure}[ht!]
\epsscale{1.0}
\plotone{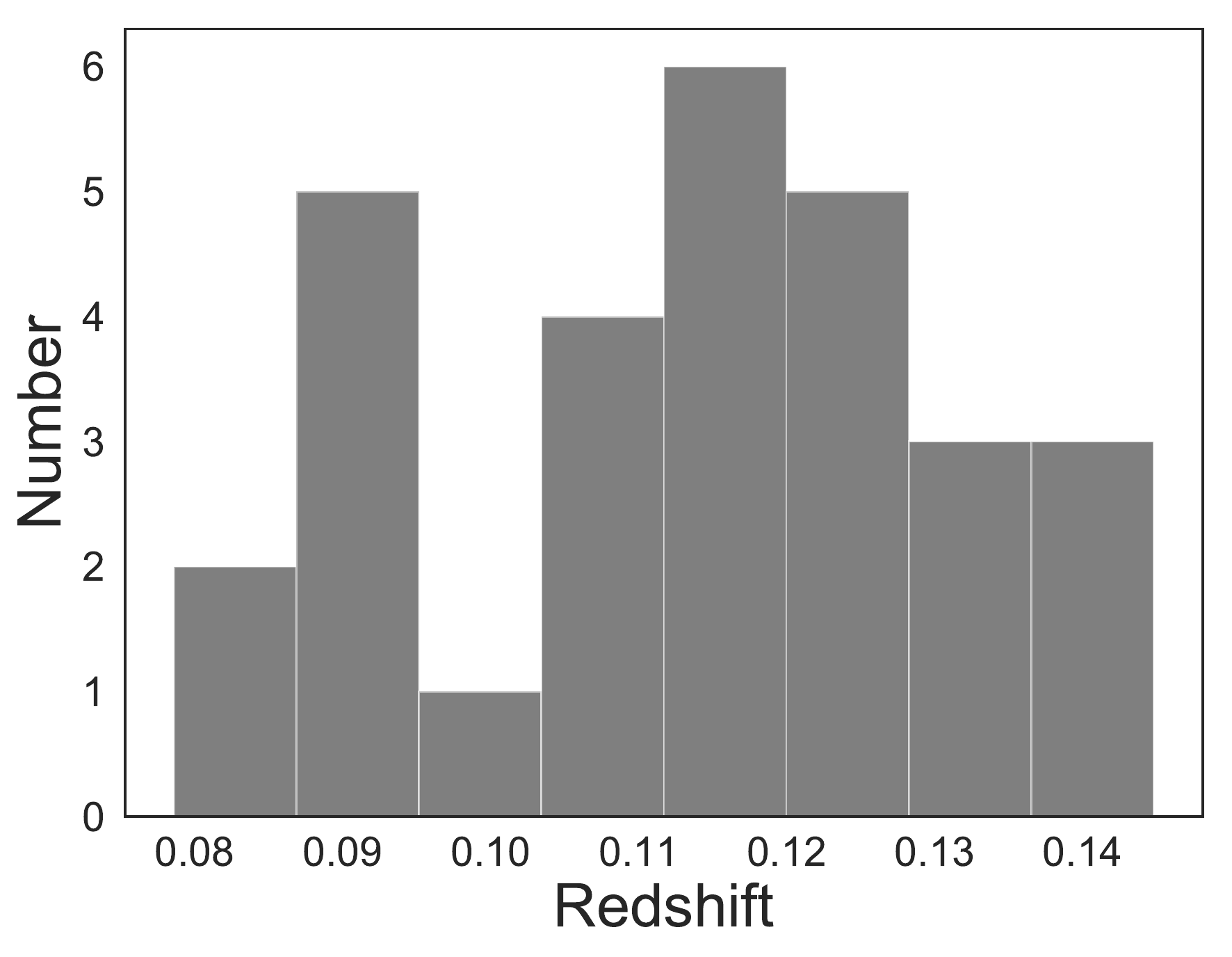}
\caption{Redshift distribution of the 29 HPS galaxies. Redshift measurements come from the HPS catalog.}.\label{fig:z_hist}
\end{figure}

After pairing line detections and consolidating duplicate lines, 29 objects remained. Thirteen of the objects have [O~II] $\lambda 3727$, [O~III] $\lambda 5007$ and H$\beta$ detections in the HPS emission line catalog. One of these 13 objects also has a [Ne~III] $\lambda 3869$ detection. Of the remaining 16, 14 were missing one line detections, and two were detected only in [O~III] $\lambda 5007$.  A list of all 29 objects and their HPS derived properties is given in Table~\ref{objs_tbl}, and a histogram of redshifts is shown in Figure~\ref{fig:z_hist}. 

\subsection{LRS2 Follow Up Observations}
\label{sec:lrs2_obs}

\begin{deluxetable*}{lcccc}
\tablecaption{LRS2 observations of HPS targets missing detections of either [O~II]$\lambda3727$,  [O~III]$\lambda5007$, or H$\beta$ \label{tab:lrs2_obs_tbl}}
\tablehead{\colhead{HPS name} & \colhead{ID} & \colhead{Date} & \colhead{Exposure Time} & \colhead{Instrument}}
\startdata
HPS030630+000128 & 44 & 2017 Dec 10 & 2414.80 & LRS2-B \\
HPS030638+000240 & 67 & 2018 Sep 23 & 2151.00 & LRS2-B \\
HPS030638+000240 & 67 & 2018 Oct 10 & 2654.95 & LRS2-B \\
HPS030651$-$000234 & 118 & 2017 Sep 22 & 2414.45 & LRS2-B \\
HPS030652+000123 & 119 & 2017 Oct 12 & 2415.25 & LRS2-B \\
HPS030655$-$000050 & 125 & 2017 Sep 18 & 1207.40 & LRS2-B \\
HPS030655$-$000050 & 125 & 2017 Sep 23 & 1207.30 & LRS2-B \\
HPS030655$-$000050 & 125 & 2017 Oct 13 & 2413.95 & LRS2-B \\
HPS030655+000213 & 129 & 2017 Oct 14 & 2413.90 & LRS2-B \\
HPS100008+021542 & 158 & 2017 Sep 19 & 2415.80 & LRS2-B \\
HPS100008+021542 & 158 & 2017 Dec 13 & 2414.80 & LRS2-B \\
HPS100021+021351 & 234 & 2019 Dec 26 & 1615.91 & LRS2-B \\
HPS100021+021351 & 234 & 2019 Jan 26 & 1816.91 & LRS2-R \\
HPS100021+021223 & 235 & 2017 Nov 29 & 2414.95 & LRS2-B \\
HPS100021+021237 & 237 & 2017 Nov 17 & 2413.95 & LRS2-B \\
HPS100028+021858 & 260 & 2017 Nov 26 & 2112.65 & LRS2-B \\
HPS100028+021858 & 260 & 2017 Dec 18 & 2414.55 & LRS2-B \\
HPS100037+021254 & 300 & 2017 Dec 23 & 1207.20 & LRS2-B \\
HPS100039+021246 & 303 & 2017 Nov 27 & 2414.65 & LRS2-B \\
HPS123632+621037 & 363 & 2017 May 13 & 2417.55 & LRS2-B \\
HPS123632+621037 & 363 & 2017 May 18 & 1208.40 & LRS2-B \\
HPS123652+621125 & 430 & 2019 Feb 02 & 1817.14 & LRS2-R \\
HPS123652+621125 & 430 & 2019 Mar 16 & 7249.40 & LRS2-R \\
HPS123652+621125 & 430 & 2019 Feb 02 & 1815.59 & LRS2-B \\
HPS123656+621420 & 438 & 2017 May 24 & 3016.35 & LRS2-B
\enddata
\end{deluxetable*}

The HPS catalog only includes lines with $5\,\sigma$ detections, so a missing entry in the catalog does not imply zero line flux. The 16 objects with missing [O~II] $\lambda 3727$, [O~III] $\lambda 5007$ or H$\beta$ detections were therefore followed up with observations from the HET Second Generation Low Resolution Spectrograph \citep[LRS2; Hill et al.\ 2019 in prep.,][]{cho16}. LRS2 is a new facility instrument consisting of two separate two-armed integral field unit spectrographs:  one (LRS2-B) covering  3700-7000~\AA\ with resolutions of $R \sim 1900$ in the UV at $R \sim 1100$ in the orange, and the other (LRS2-R) covering 6500-10500~\AA\ at $R \sim 2000$ in the red and infrared. Each unit has two spectral channels and is fed by a microlens-coupled bundle of 280, 170$\micron$-core fibers yielding a $6\arcsec \times 12\arcsec$ field of view. Each fiber has a lenslet which covers a $0\farcs 6$ hexagonal field element, creating a unity fill factor for the instrument. For a more detailed description of LRS2, see \citet{cho16}.

\begin{figure*}
\begin{center}
\includegraphics[width=.7\textwidth]{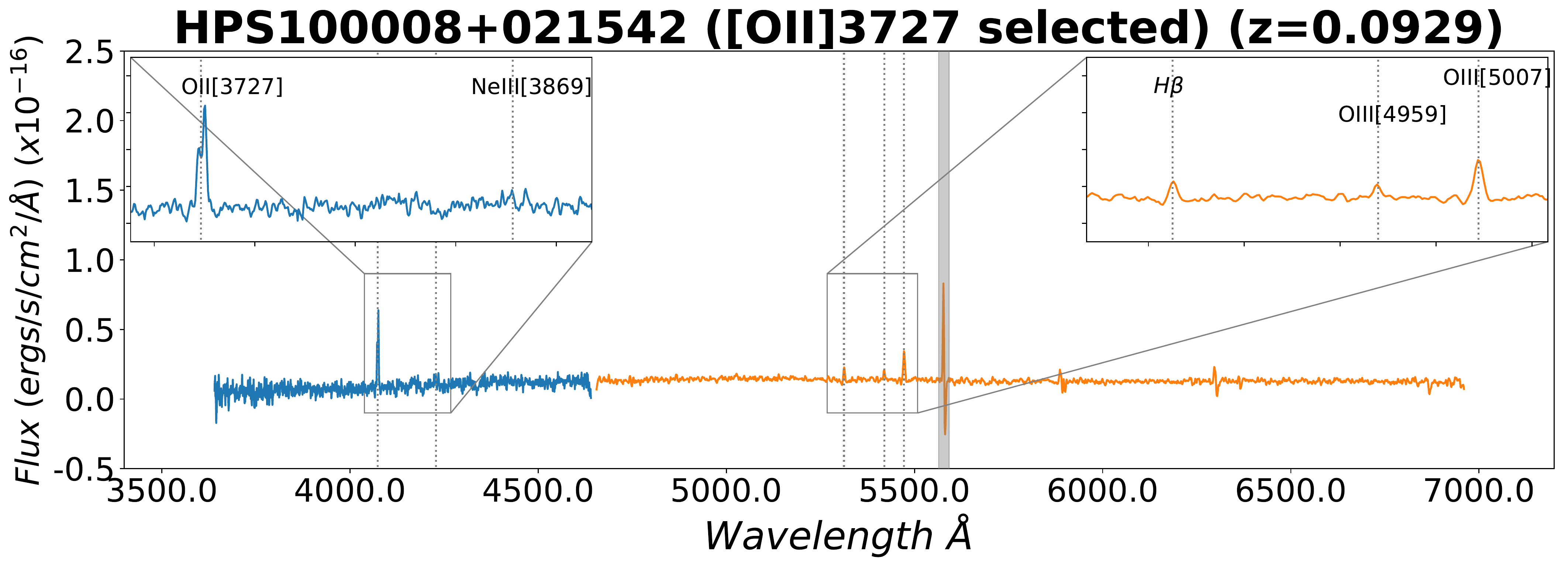}
\includegraphics[width=.7\textwidth]{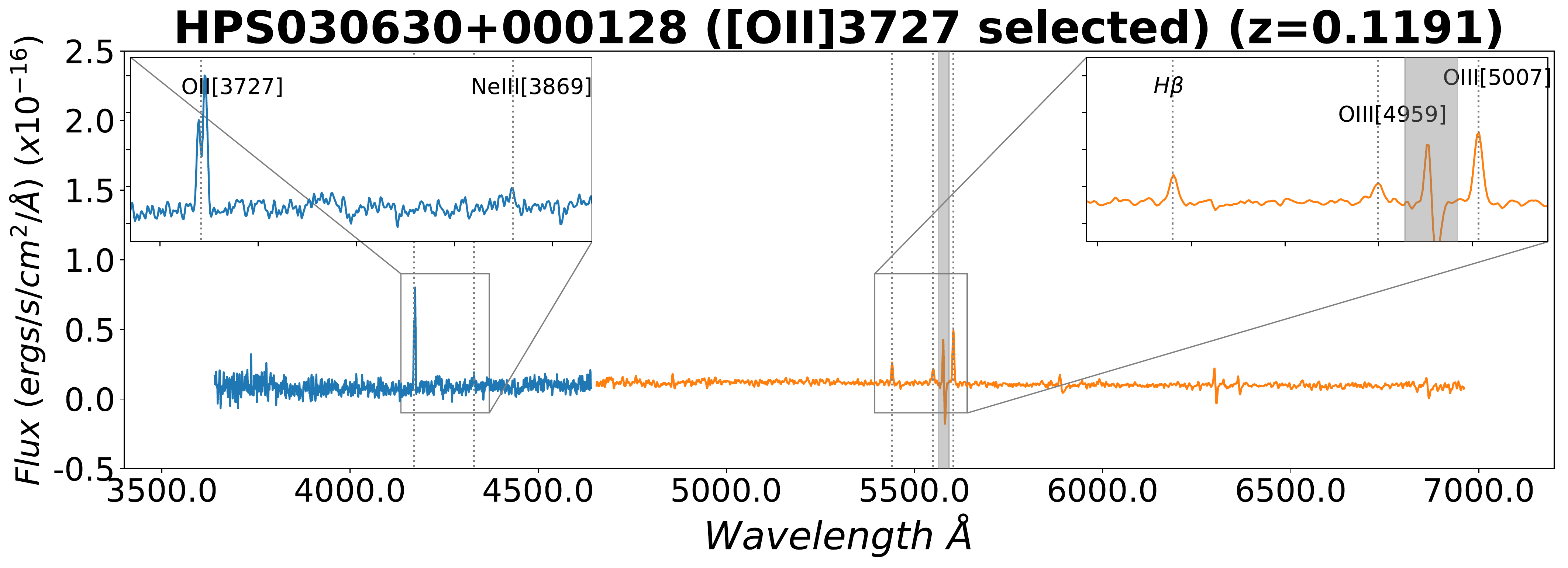}
\includegraphics[width=.7\textwidth]{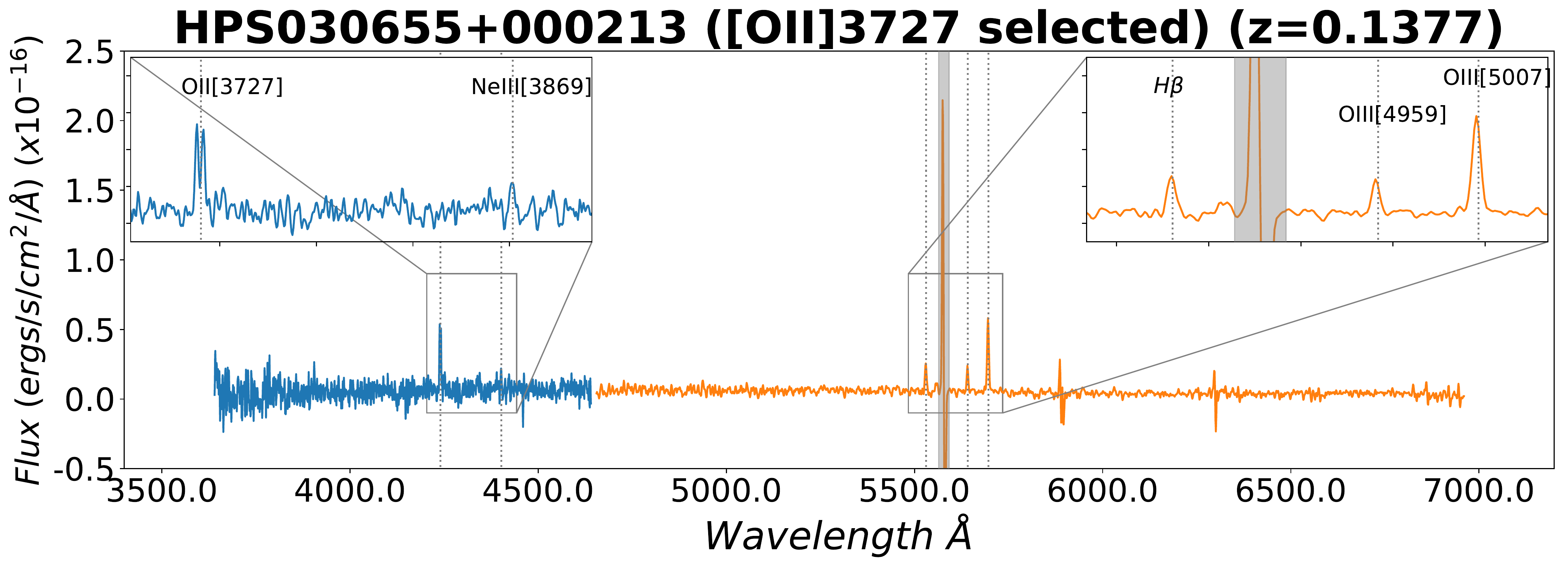}
\includegraphics[width=.7\textwidth]{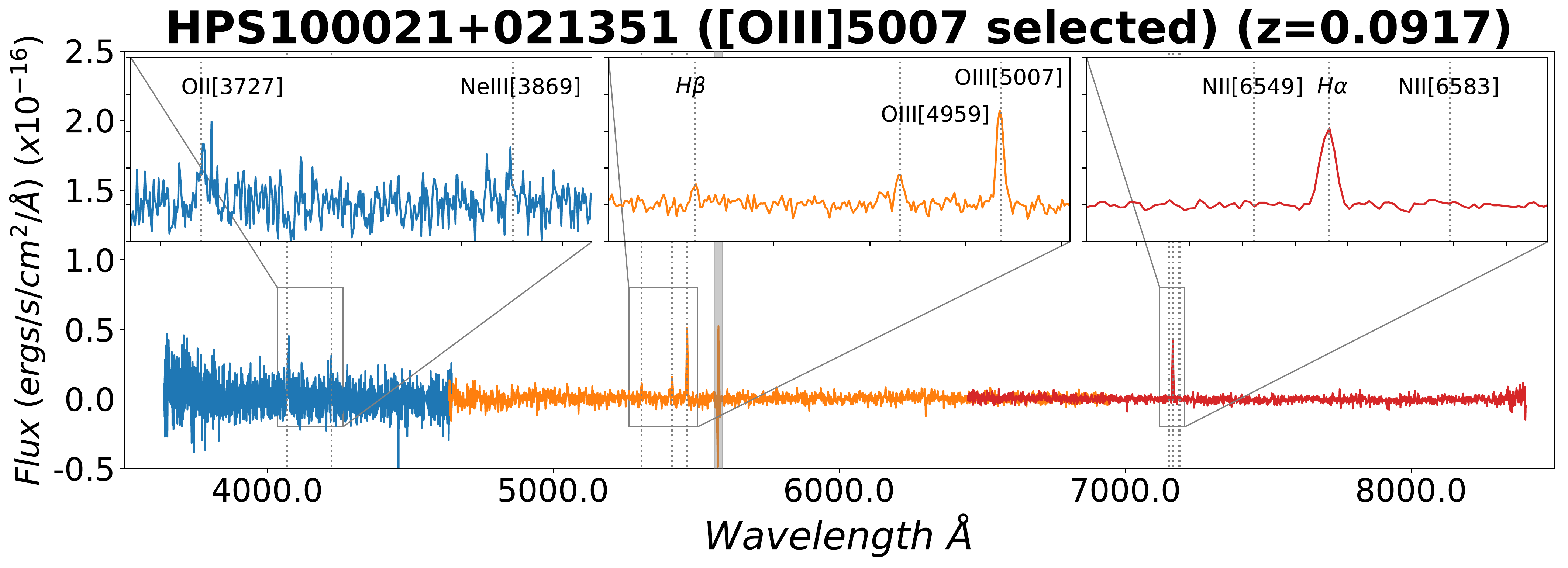}
\end{center}
\caption{Example spectra from four of the galaxies followed up with LRS2-B\null. This instrument contains two spectral channels (UV and Orange) that observe the same field simultaneously; the spectrum from the UV channel is shown in blue and the spectrum from the Orange channel is plotted in orange. The very strong [O~I] $\lambda 5577$ telluric line is shaded in grey. The measured emission lines are highlighted with grey dotted lines. The first three spectra are from objects detected in [O~II] $\lambda 3727$ in the HPS catalog and were followed up because they were missing either or both of [O~III] $\lambda 5007$ and H$\beta$. The last spectrum was only detected by HPS in [O~III] $\lambda 5007$. The [O~III] selected objects were additionally followed up with LRS2-R data, and the data from the instrument's red channel are displayed. (Data from the far-red channel are not shown because they was not used in our analysis).}
\label{fig:lrs2_spec}
\end{figure*}

Follow up data were taken with the LRS2-B unit to detect [O~III] $\lambda 5007$ in the 16 galaxies with no [O~III] entry in the HPS catalog.  In addition, two galaxies with undetected [O~II] detection were also followed up with the LRS2-R unit, in order to confirm their status as rare, low metallicity objects. Example LRS2 spectra for some of our galaxies are shown in Figure~\ref{fig:lrs2_spec}. With the HET operating on a queue based system, observations of our targets were taken over many nights from May 2018 to March of 2019 under varying conditions. A standard star was observed during each night of observations and sets of calibration data were taken as needed for reductions and flux calibration. \autoref{tab:lrs2_obs_tbl} gives an observing log for these HET observations.  \par

\subsection{LRS2 Reduction}
\label{sec:lrs2_redux}
Two independent pipelines were developed to reduce LRS2 data, the first being the LRS2 CURE-Based Quick Look Pipeline (QLP). The QLP was created to reduce and analyze LRS2 commissioning data as well as provide reduction tools to early science users. The QLP provides a Python interface for the user and calls upon routines from the Cure software package, written in C++, to run steps of the reductions. For a more detailed description of the QLP and Cure software see \autoref{sec:ql}. 

The second pipeline, Panacea (Zeimann et al.\ 2019 in prep.)  is a general integral field unit spectroscopic reduction tool tailored for the HET's two IFU-based spectrographs:  LRS2 and VIRUS\null.
The primary steps in the reduction process are bias subtraction, dark subtraction, fiber tracing, fiber wavelength evaluation, fiber extraction, fiber-to-fiber normalization,  source extraction, and flux calibration.  For more details on Panacea v0.1, which was used for the LRS2 data reductions within this paper, see the code documentation\footnote{Panacea v0.1 documentation can be found at \url{<https://github.com/grzeimann/Panacea/blob/master/README_v0.1.md>}} and Zeimann et al.\ 2019 in prep. 

As part of this study we reduced the data independently with each pipeline and achieved very consistent results for the fluxes and one sigma error bars. In the following we present results from Panacea, as this is the pipeline that will be updated and supported going forward.

\subsection{Relative Flux Calibration}
\label{sec:rel_cal}
Since the system response of LRS2 has proven to be stable over timescales longer than a several months (\citet{dav18}, Hill et al. 2019 in prep.), standard star observations taken over many nights during photometric conditions can be combined to build a system response curve with little noise. This response cureve is applied to every spectrum. However, since the standard stars are not observed simultaneously with our program targets, changes in sky transparency may affect our data.  Due to this, each reduced spectrum includes only a relative calibration and a correction for variations in the HET mirror illumination due to track position. Our approach to achieving absolutely calibrated line fluxes is discussed in \autoref{sec:abs_cal}.

\section{Measuring LRS2 Line Fluxes}
\label{sec:fluxes}
We used our reduced, combined, and relatively flux calibrated spectra to derive integrated emission-line fluxes for the [O~II] $\lambda 3726,3729$ doublet, [Ne~III] $\lambda 3869$, [O~III] $\lambda 5007$, and H$\beta$. We additionally measured H$\alpha$ and [N~II] $\lambda\lambda 6549,6583$ fluxes for the two [O~III] selected galaxies with LRS2-R spectra. Our own Bayesian emission line fitting routine was then used to fit Gaussian models which were then integrated to obtain line flux and errors (discussed in \S~\ref{sec:elf}). The lines were then absolutely calibrated to the HPS [O~II] $\lambda 3727$ or [O~III] $\lambda 5007$ measurements (discussed in \S~\ref{sec:abs_cal}). The stellar populations of our objects could not be modeled, since most of our objects have faint continua and the S/N of our spectra were, for the most part, too low for a reliable detection.  H$\beta$ equivalent widths were corrected for underlying photospheric absorption using models derived from SED fitting of photometric data (discussed in \S~\ref{sec:hb_corr}). 

\subsection{Emission Line Fitting}
\label{sec:elf}
We used our own Python based emission line fitting routine that implements a Bayesian approach to fit Gaussian models and an underlying polynomial continuum to a restricted region of the spectrum around the line(s) to be fit. 

We define the log likelihood function as $\chi^2/2$ based on the model chosen to fit the line(s). For [O~III] $\lambda 5007$, [O~III] $\lambda 4969$, and $H\beta$, we used a triple Gaussian with the ratio of [O~III] $\lambda 5007$ to 4959 fixed at 2.98 \citep{sto00}. For the two objects with red data, H$\alpha$ and [N~II]$\lambda 6549,6583$ were fit in the same way with the [N~II] lines fixed at 1:3 \citep{ost89}. This triple Gaussian  model contains four parameters: redshift, the Gaussian line width ($\sigma$), the amplitude of H$\beta$ (or H$\alpha$), and the amplitude of [O~III] $\lambda 5007$ (or [N~II] $\lambda6583$). The [O~II] $\lambda\lambda 3726,3729$ doublet was fit as a double Gaussian with parameters being the redshift, $\sigma$, and the two amplitudes. Each model also fit the continuum using a linear model with a slope and intercept.

\begin{figure*}[ht!]
\epsscale{1.4}
\plottwo{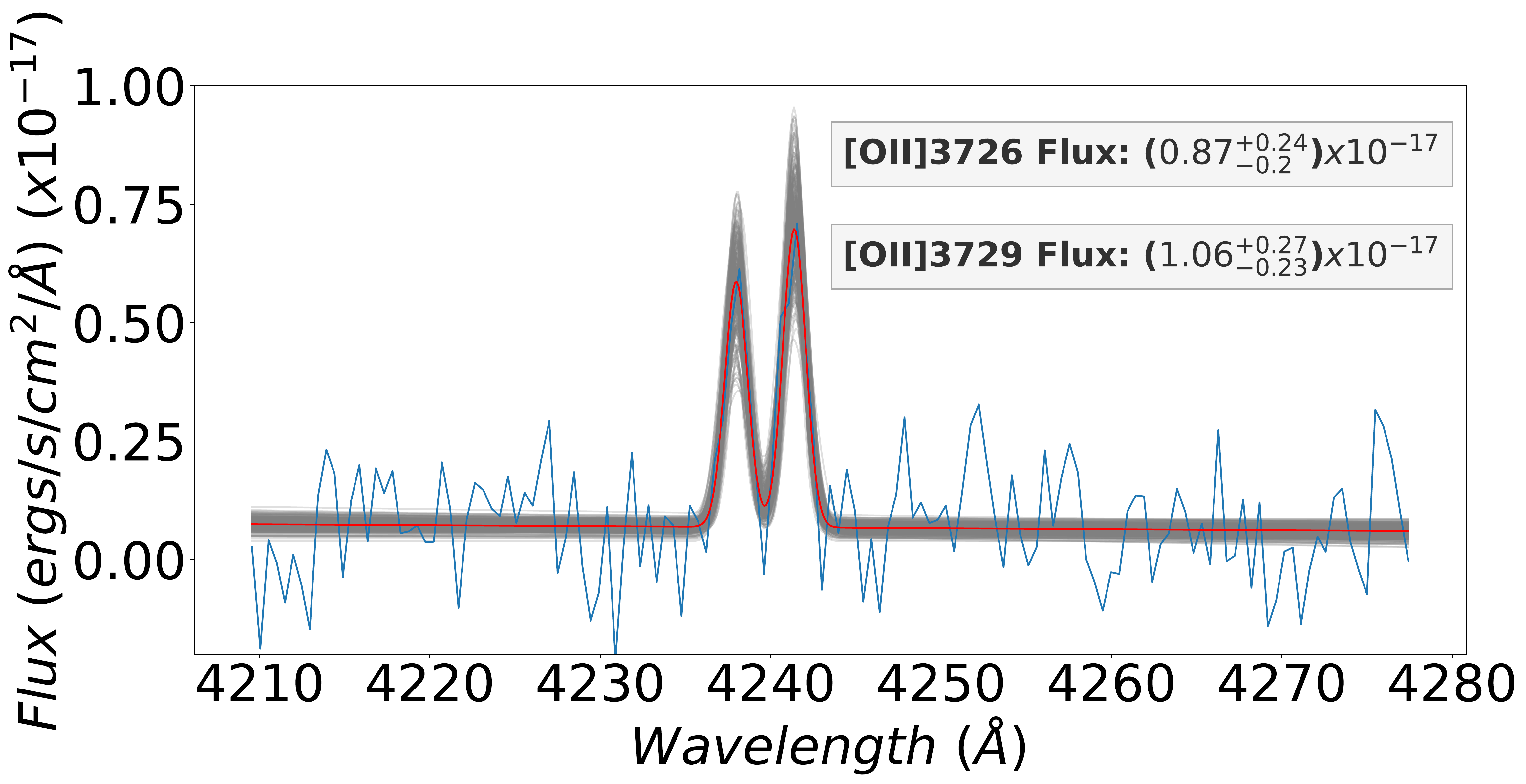}{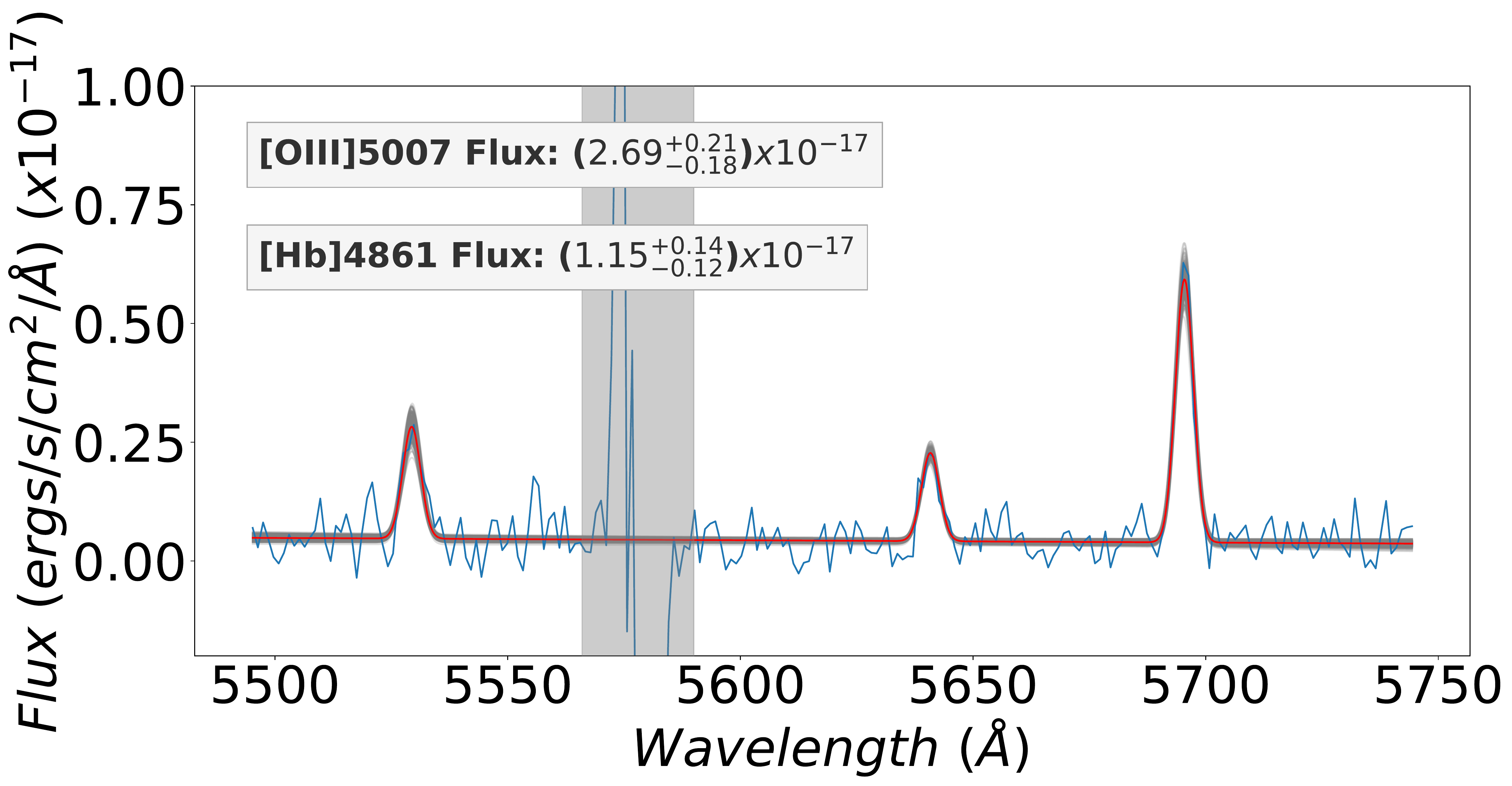}

\caption{Example output of our Bayesian emission line fitting code. The upper plot shows the fit to the [O~II] $\lambda\lambda 3726,3729$ doublet. The lower plot shows the triple Gaussian fit to the [O~III] $\lambda\lambda 4959,5007$ and H$\beta$ lines. The blue curve is the spectrum; the red is the model fit based on the results of the MCMC\null. The grey area about 5577~\AA\ represents excluded wavelengths due to the telluric [O~I] line.  The grey curves represent solutions to a hundred Markov Chains.}.\label{fig:fit_flux}
\end{figure*}


Uninformative priors were set on the model parameters (redshift, $\sigma$, line amplitude(s), and the slope and intercept of the continuum) using reasonable bounds. A lower limit of $0.5$~\AA\ were placed on the width of the emission lines based on the instrument resolution of the central wavelength of that unit. A guess for the redshift came from the HPS catalog's measured redshift of [O~II] $\lambda 3727$. Tight bounds of $\Delta z = \pm 0.001$ were placed on the redshift of the lines. The amplitude of each emission line was given the widest parameter space ($-100$ to 1000 ($ergs/s/{cm}^{2}/$\AA$~x~ {10}^{-17})$); however, a guess was placed on each by the code using the flux value of the line based on the initial guess for redshift. Even though we were fitting emission lines the amplitude was allowed to go negative since we found placing a hard limit at zero caused strange boundary errors when the amplitude was small. The slope for the line fit to the continuum had bounds of $\pm 0.01$ ($ergs/s/cm^2~x~ {10}^{-17}$), a value wide enough to accommodate for the slight gradients seen in the spectra. The $y$-intercept was probed over a wide parameter space in order to accommodate any positive value. The code uses the Python package \texttt{emcee} \citep{for13} to solve for the posterior via a Markov Chain Monte Carlo (MCMC) calculation. Example fits in both the Orange and UV LRS2 spectra are show in \autoref{fig:fit_flux}. 

Fluxes were calculated from the integral of each Gaussian, based on the the 50th percentile values of each of the Gaussian parameters output from the MCMC\null.  Integrated fluxes from both lines in the [O~II] doublet were summed and a value for the total [O~II] $\lambda 3727$ line is reported. Since the parameters were fixed for [O~III] $\lambda 5007$ and [O~III] $\lambda 4969$, a single summed [O~III] value is reported. We also fit the [Ne~III] $\lambda 3869$ line in each LRS2 spectrum; however, we did not use this line in our metallicity measurements since it was very rarely significantly detected (refer to \autoref{sec:met} for details). We consider the differences in metallicty values when using the [Ne~III] $\lambda 3869$ LRS2 fluxes in \S~\ref{sec:met}. The flux values for the lines found in our LRS2 observations of the [O~II] selected objects are presented in \autoref{tab:Flux_tbl}. The fluxes fit to the two [O~III] selected objects are presented in \autoref{tab:Flux_tbl_OIII}.

One sigma errors on each of the model parameters are output from \texttt{emcee}. Errors on the flux were calculated from these differences. 

\subsection{H$\beta$ Correction}
\label{sec:hb_corr}
The H$\beta$ line fluxes reported in the HPS catalog are not corrected for underlying H$\beta$ absorption associated with the stellar population. For most of the objects re-observed with LRS2, little continuum is detected. Instead we rely on spectral energy distribution (SED) fitting to estimate the equivalent width of the H$\beta$ absorption. The details of the SED fitting are provided in \S~\ref{sec:mass}, but briefly we use the underlying H$\beta$ absorption measured from the best-fit SED to correct H$\beta$ to a total flux.  We assume a 30 percent error on this correction based on the SED fitting code's ability to constrain the age of the galaxy with a limited filter set. The measured H$\beta$ line flux and correction are reported in \autoref{tab:Flux_tbl} and \autoref{tab:Flux_tbl_OIII}.

\subsection{Absolute Calibration of LRS2  Line Fluxes}
\label{sec:abs_cal}
 Each of the objects has either an absolutely calibrated [O~II] $\lambda 3727$ or [O~III] $\lambda 5007$ measurement in the HPS catalog.  We used these fluxes to calibrate the LRS2 spectra, by scaling our [O~II] $\lambda 3727$ measurement to match the value reported in HPS.    This ensured the line ratios seen in the LRS2 spectra were preserved.  For objects only detected in [O~III] $\lambda 5007$ in the HPS catalog, the same scaling was done but based on the [O~III] $\lambda 5007$ line. The fluxes reported in \autoref{tab:Flux_tbl} and \autoref{tab:Flux_tbl_OIII} are these normalized LRS2 fluxes.

\begin{deluxetable*}{lccccccc}
\tablecaption{Fluxes from the HETDEX Pilot Survey Catalog \citep{ada11} and LRS2 observations for the [O~II] selected galaxies. These fluxes are not corrected for reddening; the E(B-V) values used in our analysis are reported in \autoref{tab:prop_tbl}. For fluxes measured from LRS2, the errors come from the MCMC from our emission line fitting code (see \autoref{sec:elf}). These fluxes are scaled to the total [O~II] fluxes in the HPS catalog (see \autoref{sec:abs_cal}). For fluxes reported in the HPS, the errors are taken from S/N measurements listed in the catalog.  The H$\beta$ fluxes are not corrected for H$\beta$ absorption;  to correct H$\beta$ for the underlying absorption, add the correction factor reported in column H$\beta$ absorp to the H$\beta$ flux (see \autoref{sec:hb_corr}). Flux in units of ${10}^{-16}$~ergs~cm$^{-2}$~s$^{-1}$. \label{tab:Flux_tbl}}
\tablehead{\colhead{HPS Name} & \colhead{HPS ID} & \colhead{Source} & \colhead{[O~II] $\lambda 3727$} &\colhead{[Ne~III] $\lambda3870$} & \colhead{H$\beta$} & \colhead{[O~III] $\lambda5007$}  & \colhead{H$\beta$ absorp.}}
\startdata
HPS022127-043019 & 35 & HPS & 75.210$\pm$0.909 & \nodata & 29.090$\pm$0.310 & 39.230$\pm$0.375 & 8.912$\pm$2.673 \\
HPS030630+000128 & 44 & LRS2 & 4.370$\pm$0.459 & 0.371$\pm$0.300 & 1.073$\pm$0.175 & 3.055$\pm$0.273 & 0.664$\pm$0.199 \\
HPS030638+000015 & 65 & HPS & 24.380$\pm$1.037 & \nodata & 2.480$\pm$0.365 & 6.700$\pm$1.218 & 8.694$\pm$2.608 \\
HPS030638+000240 & 67 & LRS2 & 4.570$\pm$3.453 & -0.280$\pm$1.000 & 1.846$\pm$0.428 & 0.073$\pm$1.008 & 1.450$\pm$0.435 \\
HPS030649+000314 & 105 & HPS & 5.660$\pm$0.510 & \nodata & 2.370$\pm$0.354 & 5.890$\pm$0.364 & 0.498$\pm$0.149 \\
HPS030651-000234 & 118 & LRS2 & 4.030$\pm$0.791 & 0.232$\pm$0.448 & 0.706$\pm$0.117 & 1.663$\pm$0.323 & 0.680$\pm$0.204 \\
HPS030652+000123 & 119 & LRS2 & 20.760$\pm$1.991 & 1.166$\pm$1.128 & 6.770$\pm$0.306 & 13.679$\pm$0.943 & 3.366$\pm$1.010 \\
HPS030655-000050 & 125 & LRS2 & 22.390$\pm$8.311 & 2.305$\pm$4.189 & 8.760$\pm$0.450 & 14.999$\pm$2.330 & 7.120$\pm$2.136 \\
HPS030655+000213 & 129 & LRS2 & 5.080$\pm$0.973 & 1.055$\pm$0.871 & 2.835$\pm$0.269 & 7.685$\pm$0.651 & 0.593$\pm$0.178 \\
HPS030657+000139 & 138 & HPS & 9.420$\pm$0.335 & \nodata & 3.360$\pm$0.271 & 5.140$\pm$0.221 & 1.895$\pm$0.568 \\
HPS100008+021542 & 158 & LRS2 & 4.530$\pm$0.786 & 0.135$\pm$0.354 & 0.882$\pm$0.136 & 2.314$\pm$0.315 & 0.727$\pm$0.218 \\
HPS100018+021818 & 219 & HPS & 18.550$\pm$0.420 & \nodata & 4.030$\pm$0.282 & 7.100$\pm$0.252 & 2.226$\pm$0.668 \\
HPS100018+021426 & 225 & HPS & 5.460$\pm$0.250 & \nodata & 1.180$\pm$0.174 & 3.170$\pm$0.152 & 1.291$\pm$0.387 \\
HPS100021+021223 & 235 & LRS2 & 3.550$\pm$1.929 & 0.252$\pm$0.837 & 0.288$\pm$0.078 & 1.310$\pm$0.576 & 0.957$\pm$0.287 \\
HPS100021+021237 & 237 & LRS2 & 1.940$\pm$-1.153 & 0.326$\pm$-0.927 & 1.379$\pm$-0.131 & 1.179$\pm$-0.744 & 0.623$\pm$0.187 \\
HPS100028+021858 & 260 & LRS2 & 2.270$\pm$0.615 & 0.132$\pm$0.295 & 0.481$\pm$0.163 & 1.223$\pm$0.242 & 0.422$\pm$0.126 \\
HPS100032+021356 & 278 & HPS & 11.810$\pm$0.305 & \nodata & 3.680$\pm$0.201 & 15.240$\pm$0.230 & 2.086$\pm$0.626 \\
HPS100037+021254 & 300 & LRS2 & 4.180$\pm$0.495 & 0.170$\pm$0.243 & 0.778$\pm$0.131 & 1.762$\pm$0.203 & 1.670$\pm$0.501 \\
HPS100039+021246 & 303 & LRS2 & 3.110$\pm$0.541 & 0.570$\pm$0.380 & 1.627$\pm$0.255 & 3.773$\pm$0.507 & 0.407$\pm$0.122 \\
HPS100045+021823 & 326 & HPS & 7.580$\pm$0.387 & \nodata & 1.940$\pm$0.269 & 3.670$\pm$0.240 & 1.339$\pm$0.402 \\
HPS123632+621037 & 363 & LRS2 & 6.180$\pm$1.389 & -0.135$\pm$0.625 & 0.842$\pm$0.252 & 1.569$\pm$0.465 & 0.799$\pm$0.240 \\
HPS123636+621135 & 375 & HPS & 40.770$\pm$0.783 & \nodata & 14.130$\pm$0.375 & 19.270$\pm$0.326 & 9.521$\pm$2.856 \\
HPS123641+621131 & 386 & HPS & 15.340$\pm$0.496 & \nodata & 4.030$\pm$0.213 & 7.300$\pm$0.285 & 2.301$\pm$0.690 \\
HPS123648+621426 & 413 & HPS & 19.770$\pm$0.293 & 3.290$\pm$0.374 & 4.360$\pm$0.172 & 6.730$\pm$0.186 & 3.736$\pm$1.121 \\
HPS123656+621420 & 438 & LRS2 & 3.040$\pm$0.721 & -0.102$\pm$0.275 & 0.574$\pm$0.133 & 1.268$\pm$0.220 & 0.703$\pm$0.211 \\
HPS123659+621404 & 449 & HPS & 6.590$\pm$0.414 & \nodata & 1.720$\pm$0.215 & 4.890$\pm$0.223 & 1.443$\pm$0.433 \\
HPS123702+621123 & 458 & HPS & 11.430$\pm$0.257 & \nodata & 2.670$\pm$0.161 & 3.050$\pm$0.153 & 2.503$\pm$0.751
\enddata
\end{deluxetable*}

\begin{deluxetable*}{lccccccccc}
\tablecaption{Fluxes measured from LRS2 observations for the [O~III] selected galaxies. The fluxes in this table are not corrected for reddening; the E(B-V) values used in our analysis are reported in \autoref{tab:prop_tbl}. The flux errors come from the MCMC from our emission line fitting code (see \autoref{sec:elf}). These fluxes are absolutely calibrated using their [O~III] measurements in the HPS catalog (see \autoref{sec:abs_cal}). The H$\beta$ fluxes reported are not corrected for H$\beta$ absorption.  To correct H$\beta$ add the correction factor reported in column H$\beta$ absorp to the H$\beta$ flux. Flux in units of ${10}^{-16}$~ergs~cm$^{-2}$~s$^{-1}$. \label{tab:Flux_tbl_OIII}}
\tablehead{\colhead{HPS Name} & \colhead{HPS ID} & \colhead{Source} & \colhead{[O~II] $\lambda 3727$} &\colhead{[Ne~III] $\lambda3870$} & \colhead{H$\beta$} & \colhead{[O~III] $\lambda5007$}  & \colhead{H$\alpha$} & \colhead{[N~II] $\lambda6583$} & \colhead{H$\beta$ absorp.}}
\startdata
HPS100021+021351 & 234 & LRS2 & 1.013$\pm$0.687 & 0.449$\pm$0.391 & 0.392$\pm$0.188 & 2.440$\pm$0.274 & 1.428$\pm$0.152 & 0.074$\pm$0.058 & 0.120$\pm$0.036 \\
HPS123652+621125 & 430 & LRS2 & 0.902$\pm$0.483 & 0.009$\pm$0.182 & 0.143$\pm$0.093 & 0.660$\pm$0.206 & 1.208$\pm$0.127 & 0.015$\pm$0.044 & 0.300$\pm$0.090
\enddata
\end{deluxetable*}

\section{Metallicity Measures}
\label{sec:met}

For most extragalactic applications, metallicity estimates are performed by comparing the strengths of strong, collisionally-excited forbidden lines (CELs) to the strengths of the recombination lines (RLs) of hydrogen. Since the lines of oxygen are generally strong and available in more than one ionization state, it is this elemental abundance, relative to hydrogen that is most often used when quoting emission-line metallicities. 

The issue for such metallicity measurements is that the strengths of CELs are critically dependent on a number of nebular properties, such as the electron temperature ($T_e$), the electron density ($n_e$), the ionization parameter, and the hardness of the ionizing spectrum.  If one can detect faint, temperature-sensitive features such as the auroral line of doubly ionized oxygen ([O~III] $\lambda 4363$) resolve nebular density indicators, such as the [O~II] doublet $\lambda\lambda 3726,3729$, and constrain the shape of ionizing spectrum via the recombination lines of He~I ($\lambda 5876)$ and He~II ($\lambda 4686$), one can fix enough of these critical parameters and make a ``direct'' measurement of oxygen that is relatively insensitive to the remaining unknowns.  Unfortunately, for many applications, these diagnostics are unavailable:  either the electron temperature is too low to create much [O~III] $\lambda 4363$, the exciting spectrum too soft to ionize (or doubly ionize) helium, or the data have too low a resolution to constrain the density via the [O~II] $\lambda 3727$ doublet.

To help overcome these barriers, alternative methods have been developed to obtain metallicity estimates solely from the strongest emission lines in a spectrum.  Strong-line methods have the obvious advantage that, since the measurements are relatively straightforward, one can use them to obtain metallicities for large numbers of galaxies.  The disadvantage, of course, is that, given the large number of parameters that affect the strengths of CELs, an observed set of line ratios may be generated in any number of ways.  Consequently, strong-line methods must be carefully calibrated using direct metallicity measurements in galaxy samples similar to those in the systems being targeted. Such calibrations may be problematic, especially in regions of parameter space where direct measurements are difficult to come by, such as in high-metallicity systems.


For this work, we adopt one of the most commonly used sets of strong-line metallicity calibrations --- those derived by \citet{mai08} for local ($z \sim 0$) galaxies.  In the low metallicity regime (12+$\log$({\rm O/H}) $<$ 8.3) \citet{mai08} used a sample of $\sim 300$ galaxies with direct abundances derived from \citet{nag06}.  At higher abundances, a sample of $\sim 22,000$ galaxies from SDSS DR4 with log([N~II] $\lambda 6583, 6548$/[O~II] $\lambda 3727) >  1.2$ were used to calibrate the relation via photoionization models from \citet{kew02}. These high and low metallicity relations were then combined and fit with polynomials to give approximate relations between oxygen abundance and the  strong emission-line ratios.


\begin{figure*}[ht!]
\epsscale{1.0}
\plotone{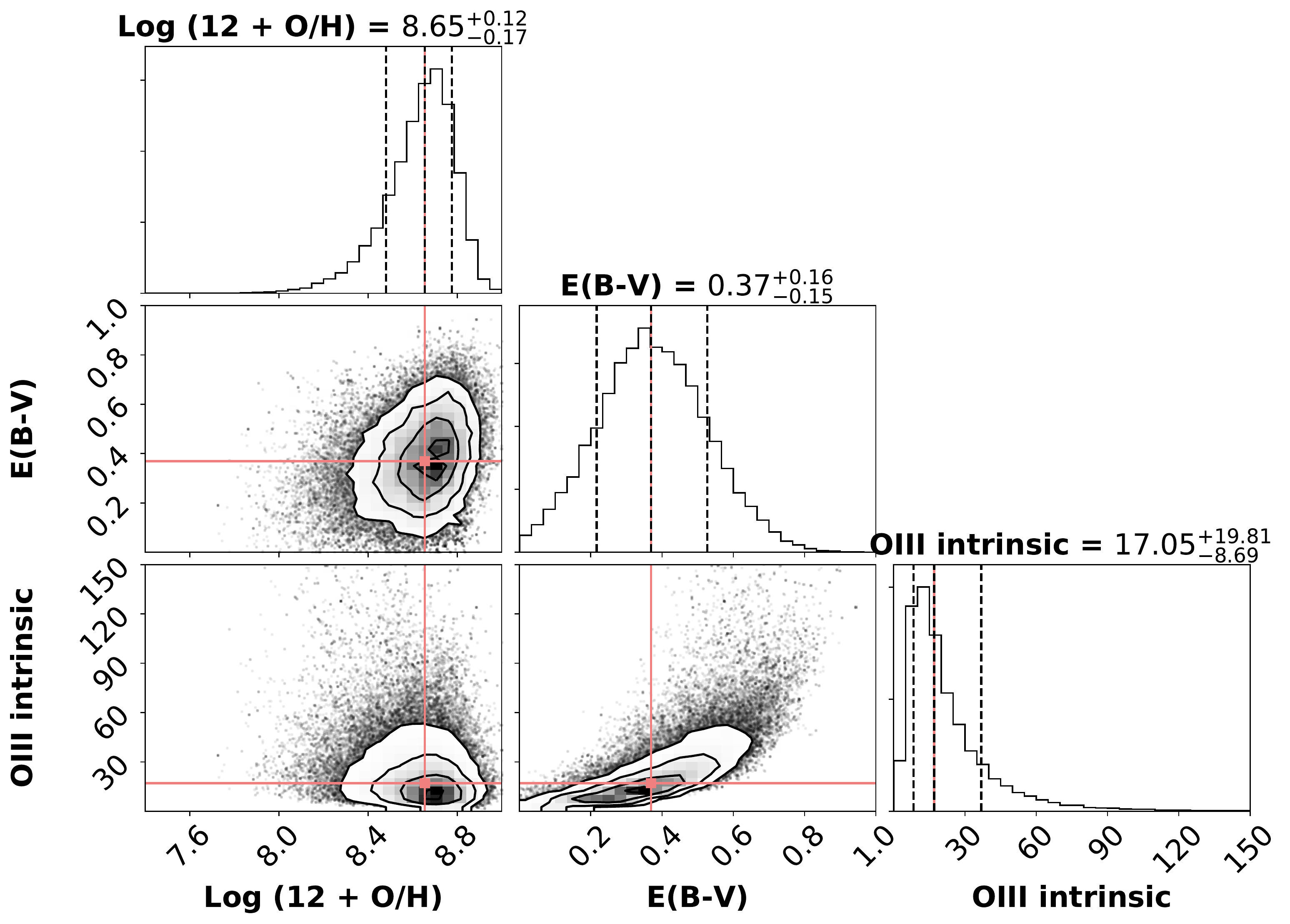}
\caption{Example metallicity fit for one galaxy from our MCMC analysis. The inner panels on the triangle plot show the posterior distribution for the three variables, metallicity (12 + log(O/H)), reddening (E(B-V)), and [O~III] intrinsic, fit by the MCMC. [O~III] intrinsic is the flux of [O~III] $\lambda$5007 corrected for reddening in units of ${10}^{-16}$~ergs~cm$^{-2}$~s$^{-1}$. The crosses represent the 50th percentile solution. The outer panels show the individual distributions for a each of the three variables. It is apparent from the metallicity vs. the E(B-V) panel that the metallicity is well constrained while the reddening is not.} \label{fig:met_results}
\end{figure*}

Since the HPS data do not extend longward of 5800~\AA\, the metallicity relations for  most of the galaxies studied here are restricted to those involving oxygen.  The most famous of these is the \citet{pag79} R23 relation
\renewcommand{\thefootnote}{\alph{footnote}}

\begin{equation}
R23\footnote{\citet{pil05, nag06}.}=\frac{{\rm [O~III]}\lambda5007+{\rm [O~III]}\lambda4959+{\rm [O~II]}\lambda3727}{{\rm H}\beta},
\end{equation}
Since metal-poor stars generally have less opacity and higher temperatures than their metal-rich counterparts, we can also take advantage of the resultant correlation between metal-abundance and ionization state
\begin{equation}
{\rm O32}\footnote{\citet{kew02, nag06, bia18}.}=\frac{{\rm [O~III]}\lambda5007}{{\rm [O~II}]\lambda3727}
\end{equation}



Finally, for the two [O~III] selected objects where we have LRS2 follow-up spectra, we can also use the reddening-independent ratio 

\begin{equation}
{\rm N2}\footnote{\citet{den02, pet04, nag06}.} = \frac{\rm [N~II] \lambda6583}{{\rm H}\alpha}
\end{equation}
\par

(We note that O3H$\beta$ = [O~III] $\lambda 5007$/H$\beta$ has also been used as a metallicity indicator, but we use R23 due to the narrower spread around its fitted relation.) All of these ratios are briefly reviewed in \citet{mai19}, but for a more detailed description of each see, the footnoted citations. 

 Finally, we note that we also have [Ne~III] $\lambda 3869$ line fluxes for objects with LRS2 follow up spectra. \citet{mai08} include a metallicity calibration of the [Ne~III] $\lambda 3869$/[O~II] $\lambda 3727$ ratio. However, due to our large errors on the [Ne~III] $\lambda 3869$ line measurements, we found adding this ratio into our analysis did not help constrain the abundance estimates. We exclude the use of the [Ne~III] $\lambda 3869$ line in our analysis. 

We applied these equations using a Bayesian approach to metallicity measurements. We define a log likelihood function as $-0.5 \,{\chi}^{2}$ where ${\chi}^{2}$ is defined as \par

\begin{equation}
{\chi}^{2} = \sum_{n=1}^{{\rm \# models}}{\frac{(x_{obs} - x_{mod})^{2}}{\sqrt{{\sigma^{2}}_{obs} +{\sigma^{2}}_{mod}}}}
\end{equation}

Here $x_{obs}$ is an array containing the observed line fluxes, the distribution of which is assumed to be Gaussian, and ${\sigma^{2}}_{obs}$, is the corresponding error term for those fluxes, based either on the measurement error if observed with LRS2 or on the S/N measurement in the HPS catalog.  Since we do not measure H$\alpha$, we are unable to directly correct the observed line ratios for reddening.  Thus, we adopt the Calzetti reddening law and treat E(B-V) as a free parameter to be marginalized over in our calculations.  
Model fluxes are derived from the \citet{mai08} relations, corrected for a given metallicity. Since these relations are in terms of line ratios, [O~III] serves as a normalization and is left as a free parameter. For the [O~III]-selected galaxies that have been followed up with LRS2-R, an additional normalization parameter is added for H$\alpha$ in order to incorporate the N2 ratio. Because the LRS2-R and LRS2-B lines are measured with separate integral field units, an overlapping emission line is required to ensure that both spectra are on the same absolute scale.  Since none of our objects have such a line, the red lines are modeled with a separate normalization. For this same reason we cannot use the Balmer decrement to fix the reddening for these objects so it is left as a parameter to be fit by the MCMC\null.  

In the equation above, $x_{mod}$ represents an array of these model fluxes and ${\sigma^{2}}_{mod}$ is the dispersion. \citet{mai08} did not present the dispersion of the models but we infer from the plots a dispersion of around 10 percent of the model ratio values in log space. The summation in the ${\chi}^{2}$ is over the number of strong lines relations used together to find a solution.\par

Uninformative priors were defined to bound the metallicity and the [O~III] (and H$\alpha$) normalizations.  The prior on metallicity is constrained to be within the values of 6.5 $<$ 12+log(O/H) $<$ 10.0. The [O~III] intrinsic parameter was very loosely bounded to be a positive value. The priors on E(B-V) were defined by measuring a distribution of E(B-V) values found for local SDSS star forming galaxies. This resulted in a prior defined by a Guassian with $\sigma$=0.165 centered at 0.295.  \par

\begin{figure*}[ht!]
\epsscale{1.0}
\plotone{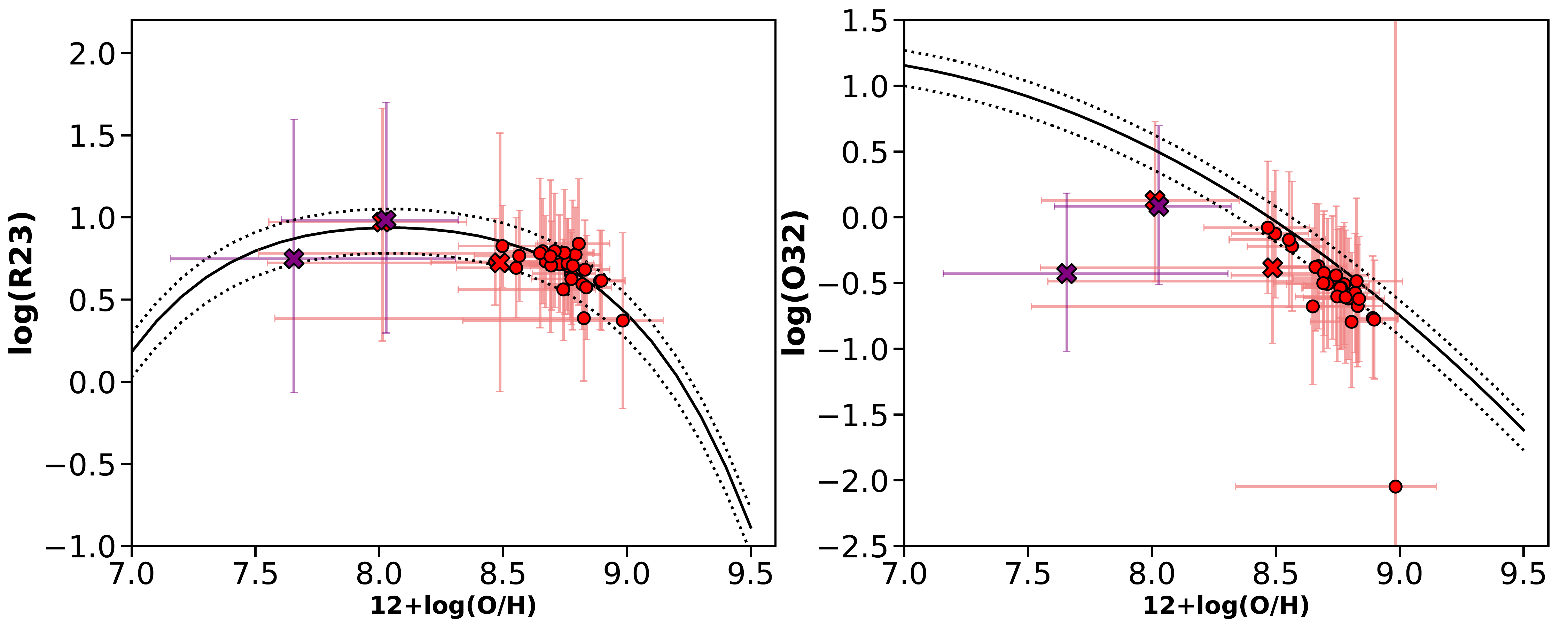}
\caption{Our metallicity estimates overplotted on the \citet{mai08} R23 and O32 relations, where the ordinate is the ratio of the observed fluxes, corrected for our MCMC-based reddening estimate and the abscissa is the metallicity derived from our MCMC analysis. The black line represents the relation from \citet{mai08} and the dotted lines represent the 10 percent error on the model. The red circles represent the [O~II] selected galaxies in the sample; the crosses denote the [O~III] selected galaxies.  For the latter galaxies, the red crosses are the metallicities measured using just R23 and O32, while the purple crosses add the red lines and N2 model into the measurement. Note that the crosses for each [O~III] selected galaxy do not necessarily have exactly the same blue strong line ratios. This is because the fluxes are corrected for the MCMC-based reddening, which does not always produce the same value when the red lines are added into the calculation. }.\label{fig:met_mai}
\end{figure*}

The Python package \texttt{emcee} was implemented to run a MCMC to solve for the posterior. An example triangle plot output from one galaxy is shown on the left side of \autoref{fig:met_results}. This gives an example of the posterior distributions derived for the metallicity, reddening, and [O~III] intrinsic flux. An example solution is shown on the right side of \autoref{fig:met_results}. The best fit values are plotted against each other in the right panel. The contours represent one sigma confidence intervals. This plot demonstrates that the metallicity is fairly tightly constrained while the reddening is not, a result consistent with the results of \citet{mai08}

In \autoref{fig:met_mai} the best fit solutions for each galaxy are plotted over the two 
\citet{mai08} relations for oxygen abundance. The R23 and O32 ratios come from the observed line fluxes and are corrected for the reddening measured from the MCMC, with the errors in the line fluxes propagated with the MCMC uncertainties in the reddening. The metallicities are measured from the MCMC and are plotted with their resulting errors. The [O~II] selected galaxies are represented with red circles; the [O~III] selected galaxies with crosses. The red crosses represent the metallicity measured with just the R23 and O32 (blue lines), while the purple crosses are metallicities which incorporate the N2 ratio (red lines). \par 

It is apparent from \autoref{fig:met_mai} that R23 is double-valued against metallicity and the O32 ratio breaks this degeneracy. Most of the galaxies have a low O32 ratio which results in the higher metallicity solution on R23. For one of the [O~III]-selected galaxies, the metallicity measured remains unaffected by adding in the red lines because the metallicity lies at the peak value of the R23 relation. However, for the other [O~III]-selected galaxy the red lines appear to drastically decrease the measured metallicity.  However, for this particular galaxy, the [O~II] $\lambda3727$ and [O~III] $\lambda5007$ lines are not well measured lines, so although the metallicity derived using just the blue lines prefers the higher metallicity solution of R23, the error is large enough to also include the low-metallicity solution. In contrast, H$\alpha$ is well measured in the long LRS2 exposure, so the non-detection of [N~II] fixes the metallicity to the lower branch of the R23 relation with smaller error bars on the measurement. Most of the [O~II] selected galaxies have [O~II]$\lambda3727$ and [O~III]$\lambda5007$ measured sufficiently well to constrain the metallicity errors to within the range of just the high metallicity branch of R23. There is one [O~II] selected galaxy (HPS100021+021223, ID=235) that has large enough error in R23 that is spans both high and low metallicity solutions, and this raises a concern since H$\alpha$ and [N~II] cannot be used as a constraint. However, since this  galaxy with high errors was only detected in [O~II] in HPS it is very likely to have a low O32 ratio, and hence falls on the high metallicity branch of R23. It is probable that adding in the red lines to the analysis of this galaxy would confirm this the high-metallicity result.

\section{Mass Measurements}
\label{sec:mass}

Stellar masses for each of the HPS galaxies were derived from spectral energy distribution (SED) fitting with \texttt{MCSED}. \texttt{MCSED} (Bowman et al.\ 2019 in prep.) is an SED-fitting code which mates flexible stellar evolution calculations with the Markov Chain Monte Carlo algorithms of \texttt{emcee} \citep{for13}.  Included within \texttt{MCSED} are sets of versatile dust extinction and attenuation laws with varying slopes and UV bump strengths, several initial mass functions (IMFs), a prescription for continuum and PAH emission from dust, a grid of nebular continua and line emission from ionized gas, options for fixed and variable stellar metallicity, and a selection of parametric and non-parametric star formation histories. This general SED-fitting code is well-suited for surveys with both spectroscopic and photometric information.

We applied \texttt{MCSED} to our galaxy sample,
using the Padova isochrones \citep{ber94, gir00, mar08} built into the Flexible Stellar Population Systhesis (FSPS) code \citep{con09}.  For these fits, we adopted a \citet{cha03} initial mass function, a \citet{cal00} dust attenuation curve, and a star formation rate history divided into age bins of 0-100 Myrs, 100-300 Myrs, 300-1000 Myr, 1-3 Gyr, and 3-12 Gyr, each with a uniform prior.  Stellar metallicity was treated as a free parameter, and the contribution of nebular emission was computed assuming an ionization parameter of $\log U = -3$.  As we are primarily interested in the stellar masses of the galaxies, emission from dust and PAH molecules were ignored.  

Our input to \texttt{MCSED}, included both broad- and intermediate-band photometry and H$\beta$ flux estimates from LRS2 and HPS\null. For the [O~II] $\lambda 3727$ selected sample we used UV, optical and near-infrared photometry gathered in \citet{bri15}. The exact bands used for each object varied depending on which of the four HPS fields the galaxy was observed in; see Table 1 in \citet{bri15} for details on these photometric brands. The two galaxies detected in [O~III] were fit using photometry from 3DHST GOODS-N \citep{ske14}; see Table 2 in \citet{ske14} for details on these data. 

\begin{figure}[ht!]
\epsscale{1.0}
\plotone{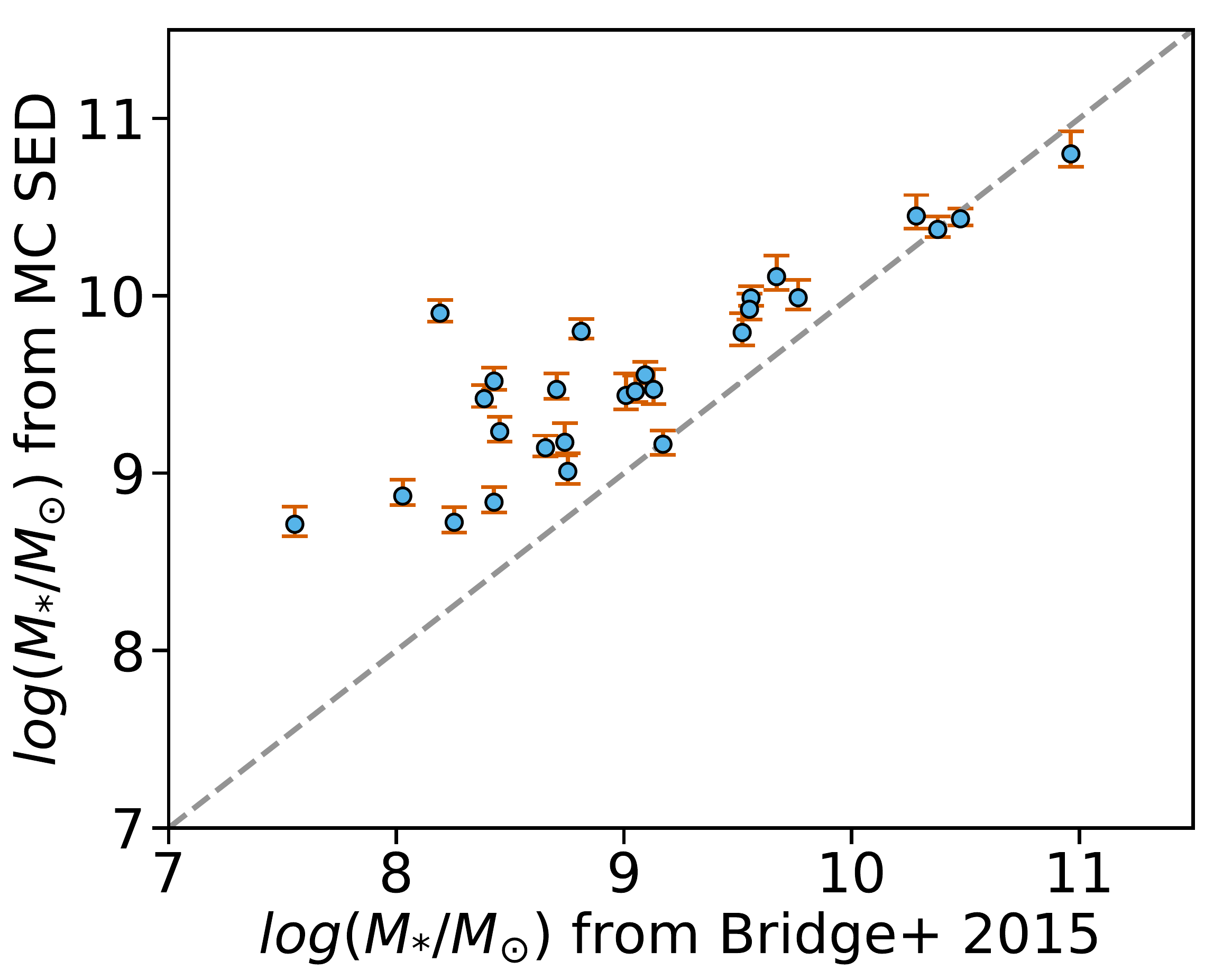}
\caption{Comparison of our \texttt{MCSED} stellar masses to those of \citet{bri15}. On the low mass end the masses derived from \texttt{MCSED} are systematically larger than the \citet{bri15} values, due to differences in the assumed star formation rate histories:  \texttt{MCSED} gives more freedom to assign higher star formation rates at older ages.  As a result, at the lower mass end, there is a growing discrepancy between the two estimates.}\label{fig:mass_compare}
\end{figure}

\autoref{fig:mass_compare} compares the
\texttt{MCSED} stellar mass estimates for our 27 [O~II] $\lambda 3727$-detected galaxies with 
those derived by \citet{bri15}.  There is a strong correlation between the two measurements, but a systematic offset as a function of stellar mass (\autoref{fig:mass_compare}).   This is not surprising:  the two codes use different star formation histories and different single stellar population models, and \citet{bri15} employed a fixed stellar metallicity while we leave ours free.  As this is a pilot survey project for the larger HETDEX survey sample, we proceed with the \texttt{MCSED} stellar masses, as this tool will be the one used future analysis. 

 The left panel of \autoref{fig:mass_met_hist} compares the distribution of masses derived for our sample to that for typical star forming galaxies in SDSS. This figure illustrates that the distribution for our sample peaks at slightly lower mass than that of SDSS. \par

\section{Mass - Metallicity Relation}
\label{sec:MZR}

\begin{figure*}[ht!]
\epsscale{1.1}
\plottwo{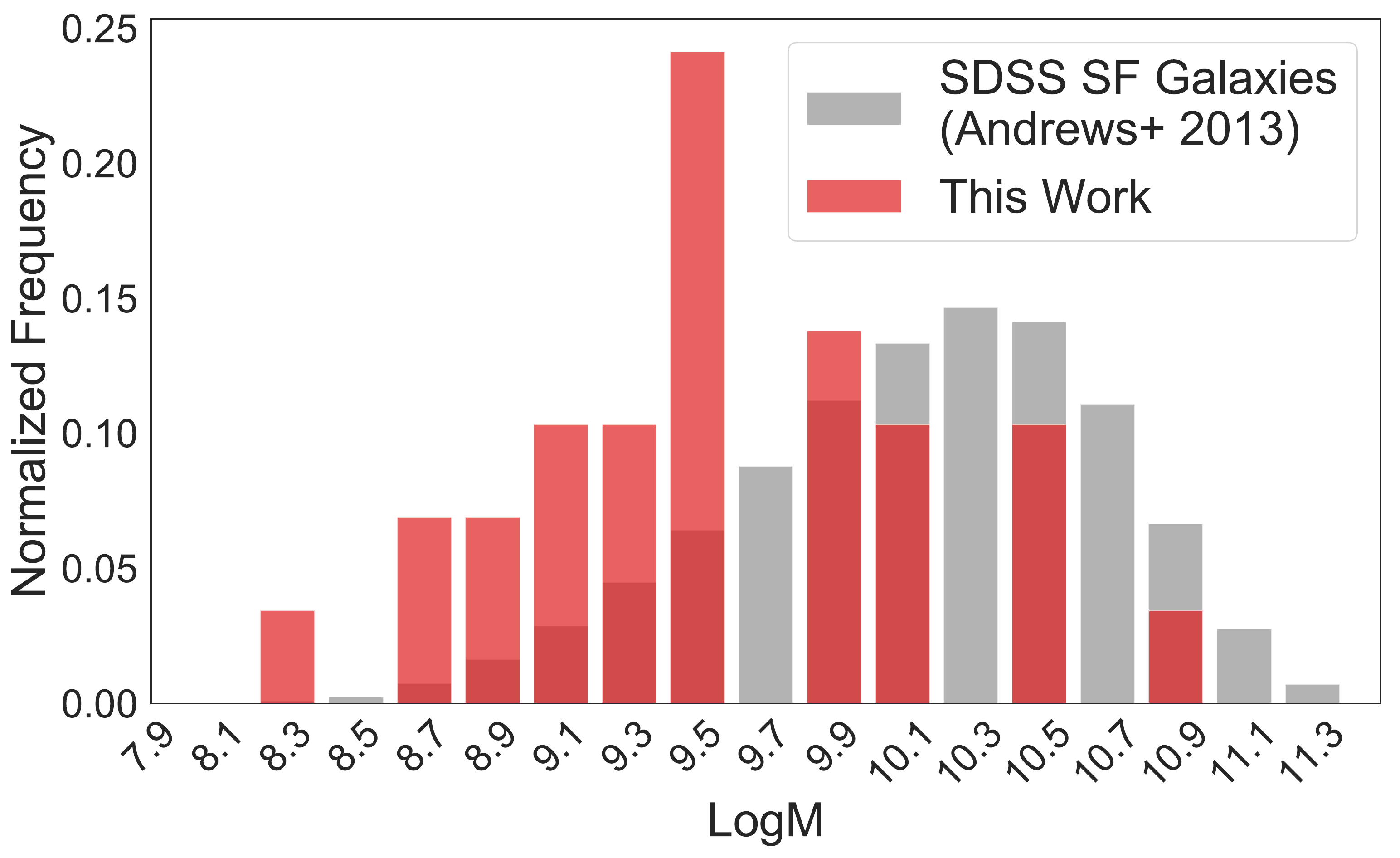}{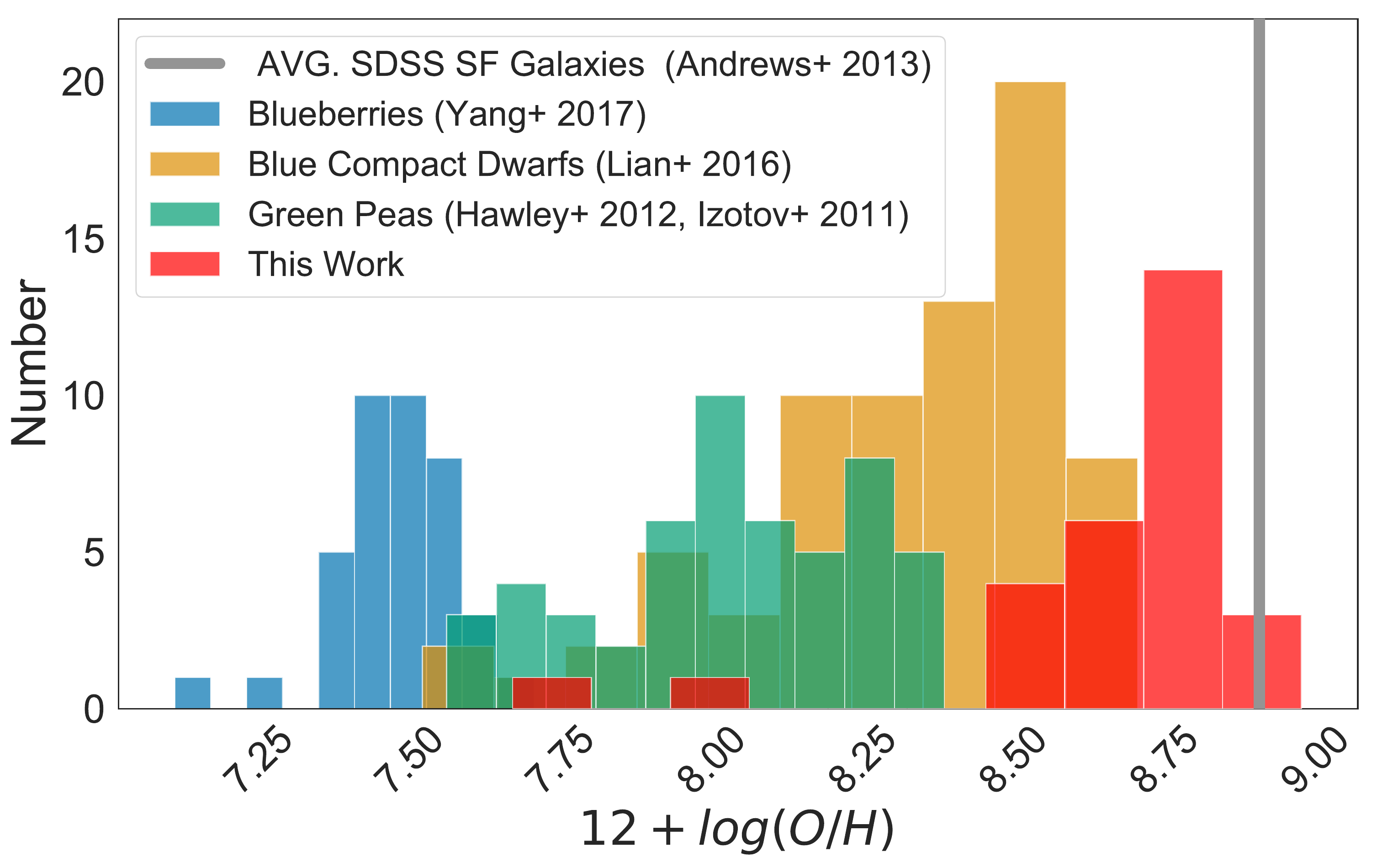}
\caption{On the left is a comparison of our sample's normalized mass distribution with that of the \citet{and13} SDSS sample. On the right is the metallicity distribution of the 29 HPS galaxies (in red).  
For comparison, three comparison samples of star forming populations measured with the same method are also plotted. The grey line represents the average metallicity (also measured with the same method) of the SDSS star forming galaxies of \citet{and13}. }.\label{fig:mass_met_hist}
\end{figure*}

\begin{figure*}
\begin{center}
\includegraphics[width=0.8\textwidth]{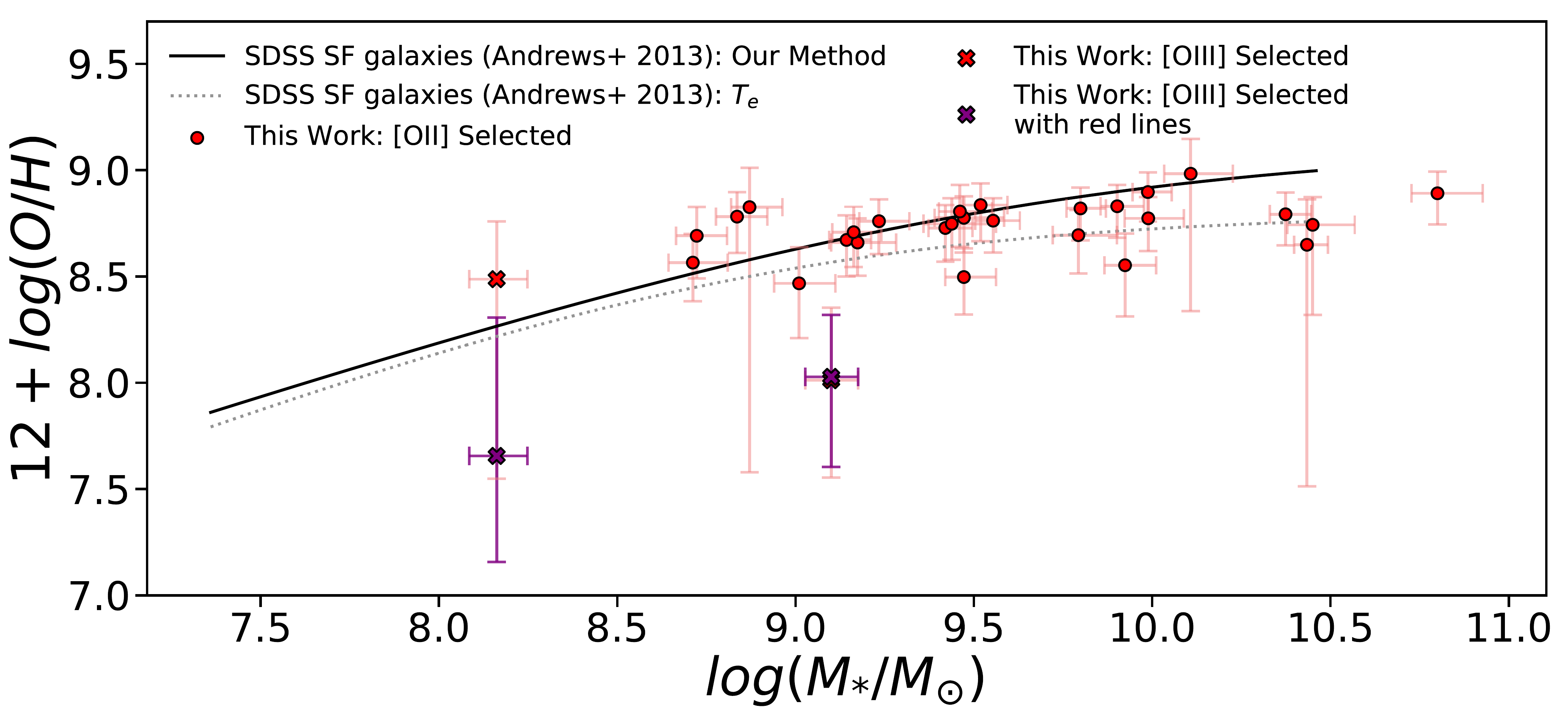}
\includegraphics[width=0.9\textwidth]{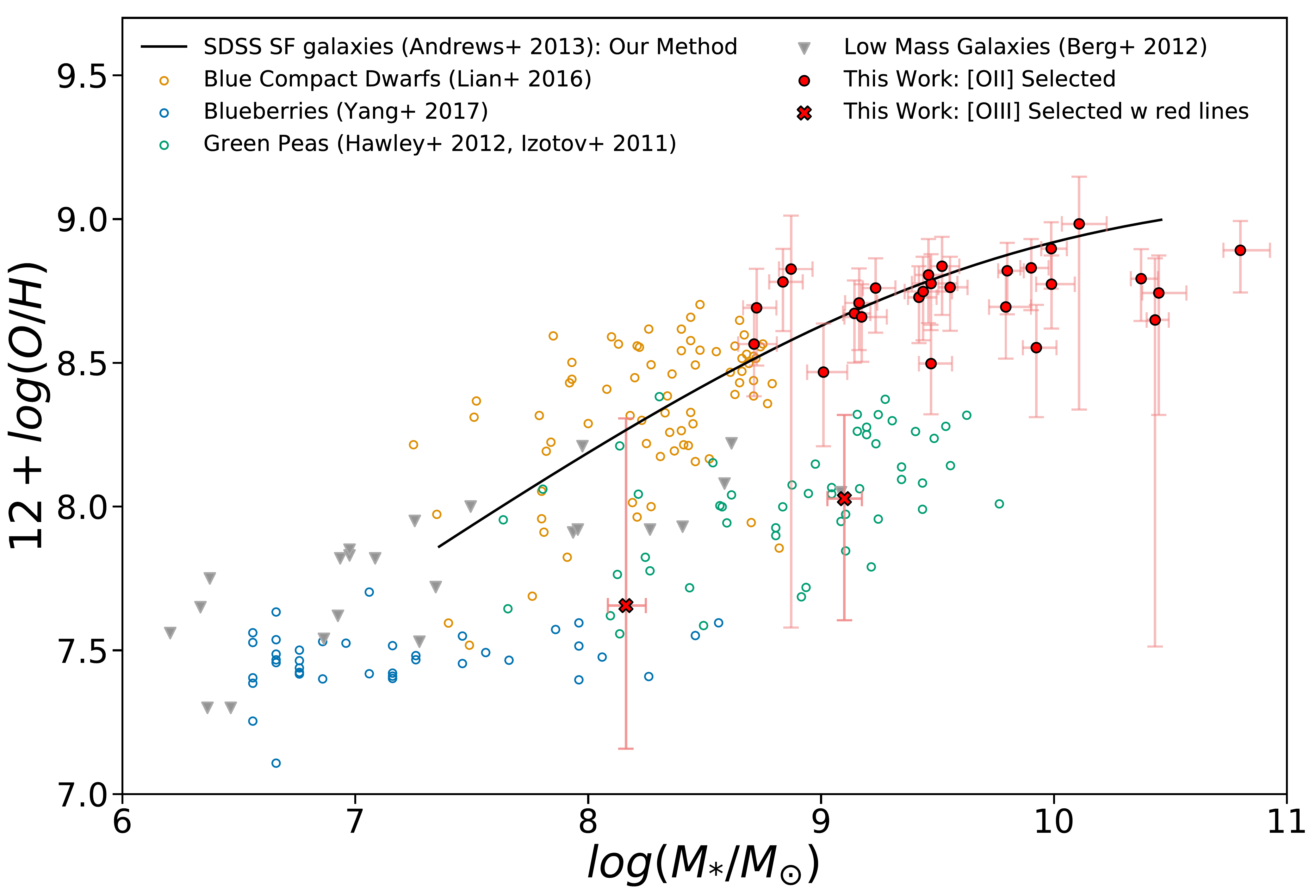}
\end{center}
\caption{ The mass-metallicity relation for our sample of 29 HPS galaxies.  In the upper panel, the dotted line shows the \citet{and13} curve derived for SDSS star forming galaxies using the direct method, while the solid line is the curve derived from \citet{and13} fluxes using our method.  The differences are mainly at the high-metallicity end, where the \cite{mai08} models are calibrated using photoionzation models.  The red circles represent the [O~II] selected galaxies, the red crosses show the [O~III] selected galaxies, and the purple crosses show metallicities derived by additionally including the red lines and the N2 model. In the lower panel the red circles and crosses  represent the [O~II] and [O~III] selected galaxies, respectively. Only the metallicity values we use going forward are plotted. The solid line is the curve derived from \citet{and13} fluxes using our method. Also plotted are the masses and metallicities for comparison populations of green pea, blue compact dwarf, and blueberry galaxies. Each of these comparison populations as well as the solid SDSS curve represent by metallicities derived using our calibration. The masses for the comparison galaxies are taken from the literature and are measured with a  variety methods, but each has been corrected to a common IMF.}
\label{fig:mzr}
\end{figure*}

The derived gas-phase metallicity values and stellar masses are reported in \autoref{tab:prop_tbl}. The distribution of metallicities of the HPS objects along with comparison populations are presented in \autoref{fig:mass_met_hist}. The mass-metallicity relation (MZR) for the HPS objects and comparison populations are plotted in \autoref{fig:mzr}. Inconsistencies between metallicity measurements from various methods have been discussed extensively in the literature \citep{pei07, kew08, bre09, lop10, mou10, lop12}.   Consequently, in order to compare our gas-phase metallicities to previously published measurements, we took each study's reported emission line fluxes and re-derived their metallicities using the method, as described in \autoref{sec:met}.  Similarly, we shifted each study's stellar mass estimates to be consistent with our assumed IMF, which comes from \citet{cha03} \footnote{To correct from \citet{sal55} to \citet{cha03} we subtract 0.04 dex \citep{muz13}. To correct from \cite{kro01} to \citet{cha03} we multiply the stellar mass (in linear space) by 0.61 \citep{con09}.}.

We first plot the MZR derived from \citet{and13} using $\sim 200,000$ star forming galaxies detected in SDSS DR7.    To derive this relation, \citet{and13} adopted the stellar masses derived by \citet{kau03} \citep[which assumed a ][IMF]{kro01}, binned the galaxies by stellar mass, stacked their spectra to recover the faint [O~III] $\lambda 4363$ line, and derived the mean metallicity of each bin via the direct method.  The grey dotted line in \autoref{fig:mzr} shows the result. The black curve represents this same work, but with metallicities derived from our Bayesian strong line method.  The curves are in very close agreement in the regime where the \citet{mai08} relations, on which our method is based, are calibrated against the strong line method.  At higher metallcities (12+log(O/H) $>$ 8.3), where the \citet{mai08} relations are calibrated against photoionization models, the two curve diverge slightly.

\begin{figure*}[ht!]
\plotone{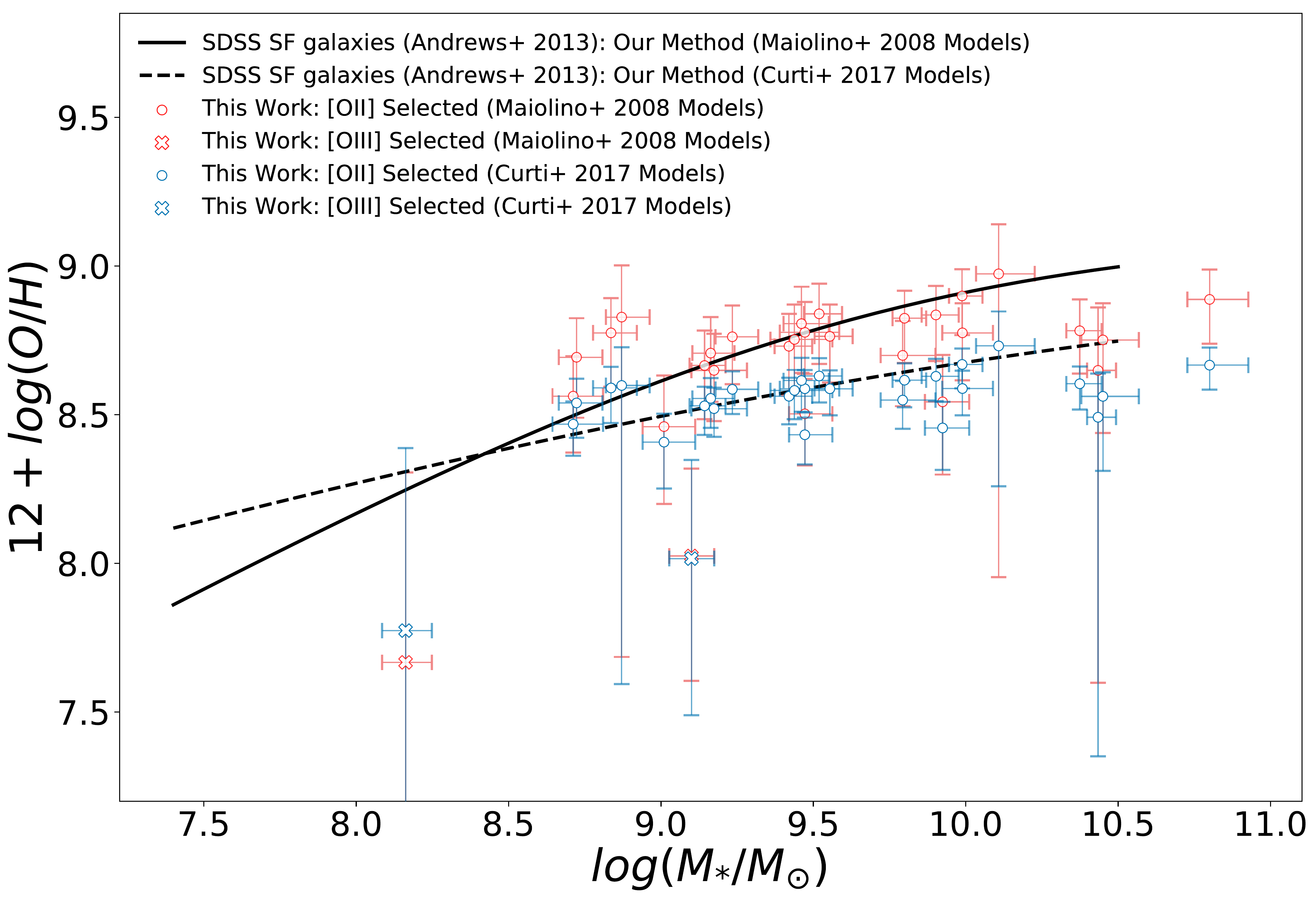}
\caption{A comparison of mass-metallicity relations derived with both the \citet{mai08} and \citet{cur17} models. The solid line is the \citet{and13} curve derived with \citet{mai08} models, and the dashed is that derived with \citet{cur17} models. The red and grey points represent our sample of galaxies measured with both \citet{mai08} and \citet{cur17} models respectivley. The [O~III] selected galaxy metallicities (denoted by crosses) are derived using the red lines (represented in purple in \autoref{fig:mzr}). Metallicities computed with the \citet{cur17} models follow the same trend as those from 
\citet{mai08}, but are systematically lower by 
$\sim 0.2$~dex.}\label{fig:mzr_compare}
\end{figure*}

More recently, \citet{cur17} were able to relate strong line ratios to direct-method abundances across the entire range of galaxy metallicity by stacking over $\sim 100,000$ galaxies in $\log$ [O~II]/H$\beta$ and $\log$ [O~III]/H$\beta$ to measure the extremely faint [O~III] $\lambda 4363$ line.  The MZR with metallicities derived from both the \citet{mai08} and \citet{cur17} relations for both our sample and the \citet{and13} SDSS galaxies are shown in \autoref{fig:mzr_compare}  (Note: for this figure, the masses are left in terms of the \citet{kro01} IMF\null.)
 Overall, the \citet{cur17} relations give lower metallicity values for all of the data points, though the difference is roughly consistent with the difference in the SDSS fit to the MZR derived from both models.  We conclude that the values change but the relative metallicities do not, so the interpretation of the results do not change depending on which of the two model sets is used. We proceed with the metallicities derived from the more well known \citet{mai08} models and consistently derived metallicities for all comparison populations. 

Typical low mass galaxies from \citet{ber12} are represented in \autoref{fig:mzr} with grey triangles. Metallicities for these galaxies are derived using the direct method, while   their masses come from 4.5~$\micron$ luminosities and K-[4.5] and B-K colors, assuming a \citet{sal55} IMF\null. This is the only comparison sample where stellar masses are not obtained from SED fitting. \citet{ber12} did compare these near-IR based stellar masses  with those found from SED fitting for a few of their galaxies and found that difference was within the error of the mass measurement.  This is also the only sample represented in \autoref{fig:mzr} that does not have metallicities derived from our method, as the data are aggregated from many studies, and a consistent set of emission-line fluxes is unavailable.  Comparisons of the \citet{and13} SDSS mass-metallicity curve derived via the direct method versus our strong-line method show that in the low metallicity regime, where these low mass galaxies typically reside, the two systems closely agree. This is unsurprising considering that at low metallicity, the strong-line relations are calibrated using the direct method.


We also include several samples of galaxies in \autoref{fig:mzr} that are photometrically selected based on their emission properties including green peas, blue compact dwarfs, and the most recent blueberry galaxies. Green peas are a class of spatially compact galaxies between $z\sim 0.112$ to $z \sim 0.360$ with a green appearance in broadband photometry due to high equivalent width of [O~III] $\lambda 5007$. \citet{car09} found 251 green pea galaxies in the SDSS release employing $r$-band color-color selection in a search for strong [O~III] $\lambda 5007$ emission. These galaxies are rare and have very low metallcity and high star formation rates for their masses, making them obvious outliers on the MZR\null. Using fluxes for these objects published in \citet{haw12} we calculated metallicity values for green peas using our method.  Masses, derived via SED fitting assuming a \citet{sal55} IMF, were taken from \cite{izo11}. They are corrected to a \citet{cha03} IMF as described in \autoref{sec:mass}.  \par 

Similar to green peas, \citet{yan17} discovered objects at lower redshifts, which they named blueberry galaxies due to the strong EW [O~III] line at bluer wavelengths.  As for the green peas, blueberry galaxies are also selected using SDSS photometry.   The objects' stellar masses, derived from SED fitting assuming a \citet{kro01} IMF that we corrected to a \citet{cha03} IMF (\autoref{sec:mass}). Their line fluxes, which we used to derive metallicities, 
were taken from \citet{yan17}.  \par

Previous to the discovery of green peas, blue compact dwarfs were a known population of galaxies selected based on their color, compact morphology, and low mass \citep{sar70, kun00}. These galaxies have strong emission in the UV (hence their blue appearance) and are actively star forming. Fluxes and masses  (derived via SED fitting using a \citet{kro01} IMF)  for a sample of blue compact dwarfs were taken from \citet{lia16} and corrected to a \citet{cha03} (\autoref{sec:mass}).  Since these galaxies do not necessarily have extreme [O~III], they occupy a region on the MZR that is more typical of low mass galaxies. \par

These three populations represent objects that are extreme in their star forming emission.  However, 
they were discovered photometrically. The green peas and blueberries were selected based on their excess in a broad photometric band containing the [O~III] line at the particular redshift of interest.  Such broadband techniques will only be sensitive to the most extreme line-emitters, since the flux of that line is spread over the width of the filter.  As the right panel of \autoref{fig:mass_met_hist} illustrates, the metallicity distribution of our spectroscopically-selected sample peaks at a higher metallicity than the extreme populations. Conversely, it also peaks at a slightly lower metallicity than the average metallicity of typical star forming galaxies in the SDSS.  \par

Blind spectroscopy in HETDEX, will allow us to find any galaxy with sufficient star formation to emit [O~II] or [O~III] above the detection limit of the survey.  This will enable us to explore whether the distribution of star formation is discrete or continuous between the galaxies that are considered typical and those that are labeled as extreme.

For the most part, the metallicities of HPS-selected [O~II] galaxies appear similar to those of SDSS photometrically-selected star forming galaxies of similar mass. This indicates that the selection techniques used in SDSS to select star forming galaxies based on their photometry is not introducing significant bias by missing galaxies. However, when we include HPS galaxies identified only in [O~III], we see they fall within the extreme regime of the green peas. With such a small sample it is not possible to determine whether there is a continuous distribution of objects between the typical SDSS star forming sysstems and the very extreme populations such as the green peas. A larger sample is needed (see \autoref{sec:hetdex}).

\begin{deluxetable*}{lccccc}
\tablecaption{Table of derived properties of our sample of 29 galaxies. Listed are the stellar masses derived from \texttt{MCSED}, the [O~II] derived (metallicity corrected) Log(SFR)s, and the metallicity and E(B-V) values found from the our MCMC emission-line measurements.  \label{tab:prop_tbl}}
\tablehead{\colhead{HPS name} & \colhead{HPS ID} & \colhead{Log$(M^{*}/M_{\odot})$} & \colhead{12+log(O/H)} & \colhead{E(B-V)} & \colhead{Log(SFR ($M_{\odot}/{\rm yr}$))}}
\startdata
HPS022127-043019 & 35 & $9.99^{+0.10}_{-0.07}$ & $8.77^{+0.10}_{-0.15}$ & $0.40^{+0.15}_{-0.15}$ & $0.90^{+0.34}_{-0.35}$ \\
HPS030630+000128 & 44 & $9.17^{+0.11}_{-0.07}$ & $8.66^{+0.11}_{-0.16}$ & $0.37^{+0.16}_{-0.16}$ & $-0.16^{+0.45}_{-0.45}$ \\
HPS030638+000015 & 65 & $10.80^{+0.13}_{-0.07}$ & $8.89^{+0.10}_{-0.15}$ & $0.34^{+0.15}_{-0.15}$ & $0.67^{+0.27}_{-0.25}$ \\
HPS030638+000240 & 67 & $10.11^{+0.12}_{-0.07}$ & $8.98^{+0.16}_{-0.65}$ & $0.42^{+0.17}_{-0.17}$ & $0.24^{+0.28}_{-0.28}$ \\
HPS030649+000314 & 105 & $8.71^{+0.10}_{-0.07}$ & $8.57^{+0.14}_{-0.18}$ & $0.40^{+0.16}_{-0.16}$ & $-0.14^{+0.53}_{-0.53}$ \\
HPS030651-000234 & 118 & $9.44^{+0.12}_{-0.08}$ & $8.75^{+0.12}_{-0.17}$ & $0.36^{+0.16}_{-0.16}$ & $-0.25^{+0.40}_{-0.39}$ \\
HPS030652+000123 & 119 & $9.79^{+0.11}_{-0.07}$ & $8.69^{+0.12}_{-0.18}$ & $0.41^{+0.15}_{-0.15}$ & $0.56^{+0.41}_{-0.41}$ \\
HPS030655-000050 & 125 & $10.45^{+0.12}_{-0.07}$ & $8.74^{+0.13}_{-0.42}$ & $0.45^{+0.16}_{-0.17}$ & $0.73^{+0.42}_{-0.43}$ \\
HPS030655+000213 & 129 & $9.01^{+0.10}_{-0.07}$ & $8.47^{+0.17}_{-0.26}$ & $0.43^{+0.16}_{-0.16}$ & $0.06^{+0.62}_{-0.61}$ \\
HPS030657+000139 & 138 & $9.47^{+0.11}_{-0.08}$ & $8.78^{+0.10}_{-0.16}$ & $0.41^{+0.15}_{-0.15}$ & $0.22^{+0.35}_{-0.34}$ \\
HPS100008+021542 & 158 & $9.16^{+0.08}_{-0.06}$ & $8.71^{+0.12}_{-0.16}$ & $0.36^{+0.16}_{-0.16}$ & $-0.37^{+0.42}_{-0.41}$ \\
HPS100018+021818 & 219 & $10.37^{+0.07}_{-0.04}$ & $8.79^{+0.10}_{-0.15}$ & $0.33^{+0.15}_{-0.15}$ & $0.52^{+0.33}_{-0.34}$ \\
HPS100018+021426 & 225 & $9.42^{+0.08}_{-0.05}$ & $8.73^{+0.11}_{-0.16}$ & $0.38^{+0.16}_{-0.15}$ & $0.05^{+0.39}_{-0.38}$ \\
HPS100021+021351 & 234 & $9.10^{+0.07}_{-0.07}$ & $8.03^{+0.29}_{-0.42}$ & $0.50^{+0.18}_{-0.17}$ & $-1.05^{+1.22}_{-1.19}$ \\
HPS100021+021223 & 235 & $10.43^{+0.06}_{-0.04}$ & $8.65^{+0.21}_{-1.14}$ & $0.41^{+0.17}_{-0.17}$ & $-0.08^{+0.55}_{-0.56}$ \\
HPS100021+021237 & 237 & $8.87^{+0.09}_{-0.05}$ & $8.83^{+0.18}_{-1.25}$ & $0.45^{+0.17}_{-0.16}$ & $-0.16^{+0.40}_{-0.40}$ \\
HPS100028+021858 & 260 & $8.72^{+0.09}_{-0.06}$ & $8.69^{+0.14}_{-0.20}$ & $0.39^{+0.17}_{-0.17}$ & $-0.49^{+0.46}_{-0.46}$ \\
HPS100032+021356 & 278 & $9.47^{+0.09}_{-0.05}$ & $8.50^{+0.14}_{-0.18}$ & $0.39^{+0.16}_{-0.16}$ & $0.24^{+0.57}_{-0.57}$ \\
HPS100037+021254 & 300 & $9.52^{+0.07}_{-0.05}$ & $8.84^{+0.10}_{-0.17}$ & $0.41^{+0.15}_{-0.15}$ & $0.03^{+0.31}_{-0.31}$ \\
HPS100039+021246 & 303 & $9.92^{+0.09}_{-0.06}$ & $8.55^{+0.15}_{-0.24}$ & $0.42^{+0.17}_{-0.17}$ & $-0.07^{+0.56}_{-0.56}$ \\
HPS100045+021823 & 326 & $9.55^{+0.07}_{-0.05}$ & $8.76^{+0.11}_{-0.15}$ & $0.36^{+0.15}_{-0.15}$ & $0.16^{+0.36}_{-0.36}$ \\
HPS123632+621037 & 363 & $9.46^{+0.09}_{-0.06}$ & $8.81^{+0.12}_{-0.17}$ & $0.33^{+0.17}_{-0.16}$ & $0.14^{+0.36}_{-0.34}$ \\
HPS123636+621135 & 375 & $9.80^{+0.07}_{-0.04}$ & $8.82^{+0.10}_{-0.15}$ & $0.42^{+0.15}_{-0.15}$ & $0.66^{+0.31}_{-0.31}$ \\
HPS123641+621131 & 386 & $9.23^{+0.09}_{-0.06}$ & $8.76^{+0.10}_{-0.15}$ & $0.36^{+0.15}_{-0.15}$ & $0.16^{+0.35}_{-0.36}$ \\
HPS123648+621426 & 413 & $9.90^{+0.07}_{-0.05}$ & $8.83^{+0.10}_{-0.15}$ & $0.35^{+0.15}_{-0.15}$ & $0.72^{+0.31}_{-0.31}$ \\
HPS123652+621125 & 430 & $8.16^{+0.09}_{-0.08}$ & $7.66^{+0.65}_{-0.50}$ & $0.49^{+0.18}_{-0.17}$ & $-1.27^{+1.52}_{-1.47}$ \\
HPS123656+621420 & 438 & $8.84^{+0.09}_{-0.06}$ & $8.78^{+0.12}_{-0.17}$ & $0.38^{+0.16}_{-0.16}$ & $-0.31^{+0.38}_{-0.37}$ \\
HPS123659+621404 & 449 & $9.14^{+0.07}_{-0.05}$ & $8.67^{+0.12}_{-0.17}$ & $0.40^{+0.15}_{-0.16}$ & $-0.18^{+0.43}_{-0.43}$ \\
HPS123702+621123 & 458 & $9.99^{+0.07}_{-0.04}$ & $8.90^{+0.09}_{-0.14}$ & $0.34^{+0.15}_{-0.15}$ & $0.52^{+0.25}_{-0.25}$
\enddata
\end{deluxetable*}

\section{Star Formation Rates}
\label{sec:sfr}

As described above, the mass-metallicity relation (MZR) of the HPS [O~II] selected galaxies is similar to that of galaxies selected photometrically by SDSS\null.  But recent papers have begun deriving a Fundamental Metallicity Relation which includes star formation rate (SFR) as a third axis  \citet{man10, man11, cre12, lar13, and13, gra16, hun16, cre18}.

\begin{figure*}[htp]
\epsscale{0.6}
\plotone{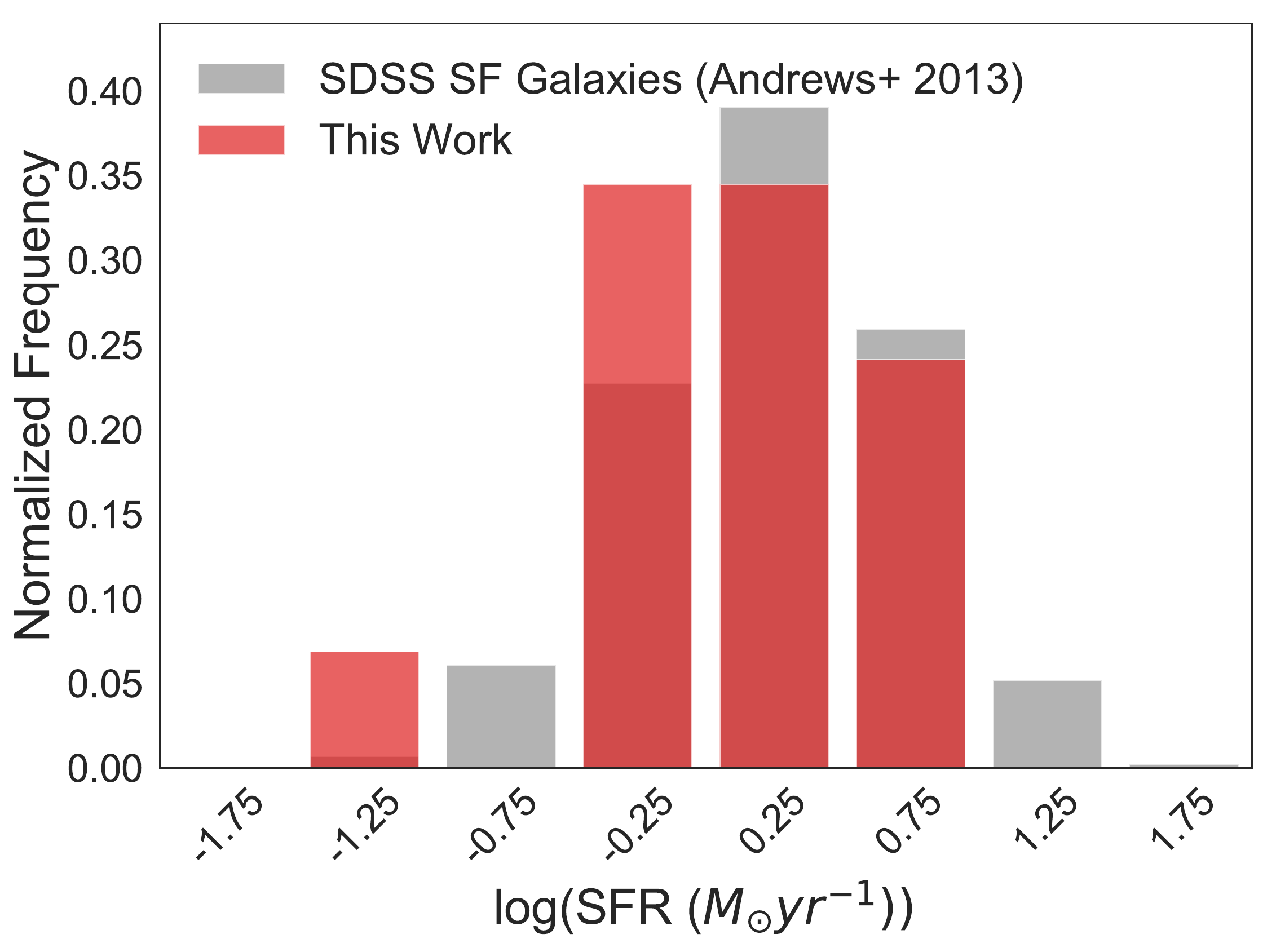}
\caption{The normalized distribution of SFRs for our sample of HPS galaxies compared to the normalized distribution of SFRs of the large sample of SDSS galaxies from \citet{and13}}.\label{fig:sfr_hist}
\end{figure*}

We derive our star formation rates (from the [O~II] line fluxes) using the \citet{kew04} relation between [O~II] luminosity, with and without a correction for metallicity. The first relation is based on the \citet{ken98} calibration with a reddening correction to the [O~II] line flux using the \citet{cal00} attenuation law and E(B-V) values derived from our metallicity code. The relation between SFR and [O~II] luminosity is  
\begin{multline}
{SFR}_{\rm{[O~II]}}(M_{\odot}\: {\rm yr}^{-1}) = 6.58\pm1.65\: \times \: 10^{-42}\\ * L({\rm [O~II]})({\rm ergs}\: {\rm s}^{-1})
\end{multline}
where $L({\rm [O~II]})$ is the reddening corrected luminosity of the [O~II] line measured for each of our objects. Since the [O~II]/H$\alpha$ ratio on which this SFR relation is based is not independent of oxygen abundance, \citet{kew04} derived an [O~II] SFR relation with an abundance correction, i.e.,
\begin{multline}
{\rm SFR}_{\rm [O~II]}(M_{\odot}\: {\rm yr}^{-1}) = \frac{7.9\: \times \: 10^{-42} L({\rm [O~II]})({\rm ergs}\: {\rm s}^{-1})}{a[12+\log({\rm O/H})]+b}
\end{multline}
The values for $a$ and $b$ are derived for different strong line indicators for 12+log(O/H) in \cite{kew04}. Of the metallicity measurement methods calibrated in \citet{kew04}, we chose the \citet{mcg91} calibration of R23 since this most closely matches our metallicity determination method. We do have additional metallicity information to distinguish between the high and low metallicity branches of the R23 relation, and because we are using a calibration that is different from that upon which the \citet{kew04} relation is based, there may be caveats to using this equation. However, in comparing the SFRs measured with and without this correction we found only small differences, with and average of 0.02 dex in SFR\null.  Notably, the galaxy most impacted by this correction is the lowest metallicity [O~III] selected galaxy, whose SFR was lowered by 0.37 dex. The metallicity-corrected SFRs are reported in \autoref{tab:prop_tbl}. The distribution of SFRs, compared with the distribution for SDSS star forming galaxies is shown in \autoref{fig:sfr_hist}. This histogram shows the SFRs for our galaxies have a similar distribution to typical star forming galaxies in SDSS. 

\begin{figure*}
\epsscale{1.2}
\plotone{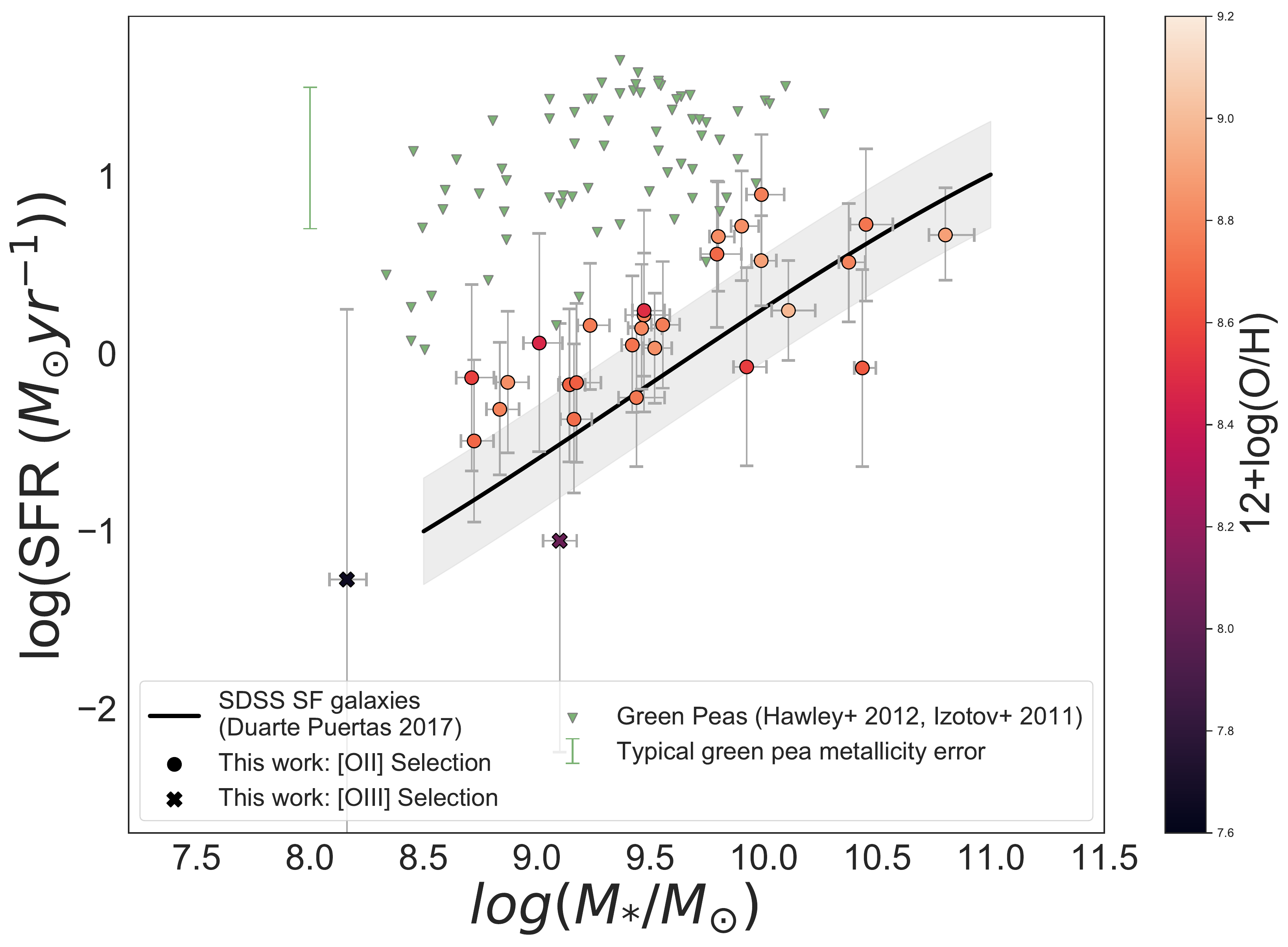}
\caption{SFR rate plotted against stellar mass for our 29 HPS galaxies. The [O~II] selected galaxies are plotted as circles, while the two [O~III] selected galaxies are shown as crosses. The points represent the metallicity corrected SFRs with the colors reflecting their measured metallicity. The black line shows the star forming main sequence for SDSS field galaxies, as derived from \citet{dua17}, with the grey region representing the 1 sigma dispersion. For comparison, the SFRs from of extreme star forming green pea galaxies (derived from their [O~II] emission from \citet{car09}) are shown as green triangles. A typical error for the green pea SFRs is shown in green in the upper left. \label{fig:sfr}}
\end{figure*}

\autoref{fig:sfr} plots SFRs derived with a metallicity correction against stellar mass. [O~II]- and [O~III]-selected sources are distinguished by circle and crosses, respectively, and each point is color-coded based on its metallicity value.  Also plotted are SFRs of the \citet{haw12} sample of green pea galaxies, using their measured [O~II] $\lambda 3727$ fluxes and the E(B-V) and metallicity derived from our metallicity code.
 Finally, for comparison, the SDSS H$\alpha$-based star-forming main sequence of \cite{dua17} is also displayed. To ensure that both the SDSS main sequence is in the same system it was computed by taking the curve presented in \cite{dua17}, translating the SFRs into H$\alpha$ fluxes  using the \citet{ken98} conversion, and then converting these fluxes into [O~II] SFRs using the expression of \citet{kew04}:
\begin{multline}
Log[SFR([O~II])] = (0.97\pm0.02)*Log[SFR(H\alpha)]\\ +(-0.03\pm0.02)
\end{multline}

From the figure it is apparent that, except for the very high-mass end, our [O~II] selected sample mostly lies slightly above the main sequence of star formation.  However, this offset is not nearly as extreme as the green pea systems.  This is not surprising:  the selection method for green peas requires them to have extreme equivalent widths, which would imply a very high rate of star formation. Our sample lies in an intermediate space between typical star formation and the extreme green peas, implying that blind spectroscopy is filling in a population that SDSS is missing: galaxies that are star forming but faint in continuum.  Interestingly, the two [O~III] selected galaxies, despite falling within the same regime as green peas in the MZR, clearly occupy a unique space. Their lower star formation is consistent with the fact that they have low metallicity for their mass. The fact that green peas are high in SFR and low in metallicity is clearly a result of their selection. The two [O~III] selected galaxies that are moderate in mass but low in metallicty and more moderate in SFR suggests that they inhabit another regime that may be missed in photometrically selected surveys and are more common than the green pea class of object. 

\section{Predictions for HETDEX Sample of Local [O~II] and [O~III] Emitters}
\label{sec:hetdex}

Based on the 29 galaxies in the emission-line flux-limited HETDEX Pilot Survey sample, we conclude that many of the HPS galaxies occupy a regime of the mass-metallicity relation (MZR) that is typical of continuum-selected galaxies, but their SFRs are generally higher than that predicted from the star forming main sequence. With such a small sample, it is interesting that we have uncovered galaxies that occupy a unique part in mass, SFR, and metallicity phase space. This may indicate that emission-line spectroscopy without pre-selection is more sensitive to finding low metallicity galaxies, and a population of star forming galaxies with faint continua. \par 

The much larger HETDEX survey will sample more than 90~deg$^2$ of sky with blind spectroscopy, making it the largest sample of galaxy spectra collected without pre-selection. The statistical power of such a large sample will allow us to better understand the systematic differences in selecting galaxies in this way compared to continuum selection, and will provide a unique understanding of the very local galaxy population. \par

The HETDEX survey will sample a much larger area of sky compared to HPS, but with slightly more limited spectral coverage on the red end (3500-5500~\AA) covering [O~III] $\lambda 5007$ at $z < 0.1$. The volume sampled by HETDEX for these galaxies is still a factor of ~680 larger than that of HPS, and reaches a fainter line flux limit.

In order to estimate the expected number of local star forming  galaxies in HETDEX, we integrated the luminosity function of HPS [O~II] $\lambda 3727$ emitting galaxies \citep{cia13} down to the flux limits of HPS and HETDEX and scaled by area. From this analyis we expect to find ~33,000 line emitting galaxies with $z < 0.1$ within the entire HETDEX survey. 

The spectroscopically-selected sample in HETDEX will be sensitive to much lower SFRs than surveys which select via broadband optical photometry. With the monochromatic flux limit of HETDEX being $3.5 \times 10^{-17}$ (ergs~cm$^{-2}$~s$^{-1}$) we will be able to detect SFRs down to $0.0015 \, M_{\odot}~$~yr$^{-1}$ at the survey's median luminosity distance and down to $0.0058 \,  M_{\odot}$~yr$^{-1}$ at $z = 0.1$. HETDEX will also be much more sensitive to fainter [O~III] $\lambda 5007$ compared to existing green pea and blueberry galaxy samples. (In the case of green peas, this factor of $10^3$.) This will allow us to build a more complete and continuous picture of the relationship between mass, metallicity, and star formation. 

\section{Summary and Conclusions}
\label{sec:conclusions}

With the HETDEX Pilot Survey Emission Line Catalog we assembled a sample of galaxies detected in either [O~II] $\lambda 3727$ or [O~III] $\lambda 5007$ at $z< 0.15$ (after which [O~III] $\lambda 5007$ is redshifted out of the spectral range of the survey). Our sample consisted of 27 galaxies detected in [O~II] $\lambda 3727$ and 2 galaxies only detected in [O~III] $\lambda 5007$. Sixteen of these galaxies were followed up with the HET LRS2 to obtain deeper spectroscopy and measure [O~II] $\lambda 3727$, H$\beta$, and [O~III] $\lambda 5007$ lines not detected in the HPS survey. We implemented a Bayesian scheme to derive metallicities from the observed strong-line ratios using the calibrations of \citet{mai08}. This scheme also allowed us to marginalize over the reddening of each galaxy, which could not be measured directly due the survey's limited wavelength coverage.  [O~II] metallicity-corrected SFRs were derived using the relation from \citet{kew04}, and sellar masses were derived using SED fitting implemented by \texttt{MCSED}. We find that the distribution of stellar masses of our sample  is skewed toward smaller values compared to that of typical star forming galaxies in SDSS (see to \autoref{fig:mass_compare}).  \par

We compare the mass-metallicity relation (MZR) of our sample of spectroscopically-selected galaxies with that of typical star forming galaxies in SDSS, and several low mass, photometrically-selected galaxies that are considered to be extreme in their [O~III] $\lambda5007$ equivalent widths.  We find that galaxies selected by [O~II] tend to follow the MZR, while the two galaxies only selected in [O~III] have lower metallicities for their mass, and occupy the same region of space as that of green peas. However, in the SFR vs.\ stellar mass diagram, we see that our sample of [O~II] selected galaxies lies in an intermediate regime between the typical star forming main sequence and the extremely star forming green peas. It is also apparent that the [O~III] selected galaxies lie along the star forming main sequence, farthest from the green peas, despite being similar in mass and metallicity. We conclude that blind spectroscopic selection is filling in part of the mass-metallicity-SFR phase space missed in photometrically selected surveys like SDSS. We are finding galaxies that are actively star forming but faint in continuum.

Our sample is complementary to SDSS in its method of selection, and fills in star forming galaxies that may be missed by photometric selection due to their faintness in the continuum. From a cross-check coordinate search of the SDSS catalog, we found that only 4 of the 29 galaxies in our sample have spectra in SDSS. With the blind spectroscopic HETDEX survey, we expect to increase the sample of such galaxies by a factor of $\sim 1000$, allowing us to better fill in this void in galaxy property phase space.

\acknowledgments

We thank the staff of the Hobby Eberly Telescope for support of our LRS2 observations as well as the LRS2 comissioning team. The LRS2 spectrograph was funded by McDonald Observatory and Department of Astronomy, University of Texas at Austin, and the Pennsylvania State University. We thank the Cynthia and George Mitchell Foundation for funding the Mitchell Spectrograph used for the HETDEX Pilot survey observations. We additionally thank the HETDEX Pilot Survey consortium for building the emission line and photometric catalogs used in this work. We thank the Institute for Gravitation and the Cosmos which is supported by the Eberly College of Science and the Office of the Senior Vice President for Research at the Pennsylvania State University. We thank Karl Gebhardt for assisting us in locating HPS data products. The supercompter Maverick hosted by the Texas Advanced Computing Center was used in the storage and reduction of our LRS2 data. This research was partially funded through McDonald Observatory. 

%

\vspace{5mm}
Facilities: \facilities{Smith(VIRUS-P), HET(LRS2), HET(VIRUS)}


\software{\texttt{astropy} \citep{ast13, ast18},  
          \texttt{emcee} \citep{for13}, 
          \texttt{Cure}, \citep{sni12, sni14}
          \texttt{LA Cosmics} \citep{van01}
          \texttt{MCSED} (Bowman et al.\ 2019 in prep.)}



\appendix
\label{appendix}

\section{LRS2 CURE-Based Quick Look Pipeline}
\label{sec:ql}

The LRS2 Cure-Based Quick Look Pipeline (QLP) was developed to provide a user friendly reduction package to get early science users of the LRS2 instrument to the point of sky subtracted, extracted 1D spectra on a uniform wavelength grid, as well as provide data products to help the user visualize their three dimensional data. The user interacts with the pipeline simply through a configuration file in which they provide the date of observation and the object name(s). The configuration file also allows the user to choose which steps of the reduction they would like to run and options for those different reduction steps. The quick look pipeline automatically finds the user's data and all the calibration data needed for reduction. Since this pipeline was written during commissioning and to reduce early shared risk science observations, the pipeline accommodates different and changing data structures, header keywords, calibration data sets, and LRS2-B CCD changes. The pipeline tracks these early changes so the user does not have to navigate instrument modifications when dealing with their data. The pipeline, which was written in Python, calls upon CURE routines to run.

The QLP is a Python wrapper that provides reduction for LRS2-B and LRS2-R\null. Each unit and night gets reduced independently since calibration data for each is different.  QLP takes advantage of Cure's statistics and fits-file handling tools to preform some steps of the basic reduction, sky subtraction and fiber extraction algorithms. 

\subsection{Cure}
\label{sec:cure_ql}

Cure, a software package in C++, contains reduction and analysis routines developed for the HETDEX survey. Cure was originally intended to provide general tools for reduction and analysis of data from any IFU instrument. However, it contains individual tools and is not in itself a reduction package. A specific build of Cure that defines LRS2 in it's configuration file exists for users. This specification defines the CCD size and allows the reduction of a single IFU channel, as opposed to the many parallel IFUs of VIRUS\null. For a more detailed description of Cure refer to \citet{sni12, sni14}.

\subsection{Calibration Data Set}
\label{sec:cals_ql}
Each night a set of calibration data are taken at the HET for both LRS2-B and LRS2-R\null.  The facility calibration unit (FCU) feeds light to each unit through one of two liquid light guides optimized for either the red or blue. The FCU output consists of a series of Fresnel lenses to mimic the rays from the 10~m HET pupil and the calibration light then passes through the wide field corrector to the HET focal surface. Light for fiber flats is fed with a laser driven light source (ldls) for LRS2-B and a QTH lamp for the LRS2-R\null.  Illumination from Hg, Cd, and FeAr line lamp is also fed to both LRS2 units from the FCU for standard arcs \citep{lee12}. A set of bias frames is also taken each night. A standard set of darks is no longer taken, as the dark count for the LRS2 CCDs is so low that dark subtraction was found to introduce noise to the data. This data set is taken once every night of LRS2 observations. 

\subsection{Basic CCD  Reduction}
\label{sec:basic_ql}

The LRS2 quick-look pipeline preforms basic reduction by first trimming and subtracting the overscan from each science and calibration frame. Cosmic Ray rejection is performed on all science frames using \texttt{cosmics.py}, which implements the L.A. Cosmic algorithm \citep{van01}. A master bias is created from bias frames through the Cure routine \texttt{meanfits}, which implements kappa sigma clipping.  The master bias is subtracted from the flats, arcs, and science frames. Dark subtraction is not performed at this stage due to the dark count being so low that this would only add more noise to the data. Each of the 4 detectors in LRS2 have two amplifiers. The amplifiers are combined by first multiplying the data from each amplifier with its respective gain value as produced in the header. The arcs and flats are each normalized by their exposure times and then again combined through Cure's \texttt{meanfits}.
	
The basic reduction provides a master trace and master arc image, which is used for finding the fiber traces for sky subtraction and fiber extraction and building the wavelength solutions. Note that flat fielding is not performed, but pixel to pixel variations are accounted for through weights derived while building the distortion solution described in \S~\ref{sec:dist_ql}. These weights are implemented in the extraction of the fibers.


\subsection{Building Distortion Solution and Wavelength Solution}
\label{sec:dist_ql}

Next, a Cure routine called \texttt{Deformer} is called to find the fibers in the images, build their profile perpendicular to the dispersion direction, and create a wavelength solution for each fiber. As with any IFU dataset, each fiber spectrum has a different curvature and profile in the image. \texttt{Deformer} uses the master trace, the master arc, and a list of arc lines to find the fiber traces and wavelength solution. Seventh order Chebychev polynomials are fit to the fiber traces and arcs to build a 2D coordinate system which transforms from CCD $(x, y)$ coordinates to $(w, f)$ coordinates. The $w$ coordinate moves along the constant wavelength direction, and $f$ moves in the fiber direction and is equal to one reference $y$ coordinate.

\texttt{Deformer} also uses an initial guess for a Gauss-Hermite expansion to fit overlapping profiles to three fibers at a time. This solution is used as an initial guess for the next set of fibers. This process is iterated until it converges on a solution. Chebychev polynomials are then fit over the variation in the Gauss-Hermite fits. This fiber model provides statistical weights, which describe how much flux from each fiber contributes to each pixel. The weights described by this model essentially provide the flat field as they are used in the fibers extraction in later Cure routines.

\subsection{Sky Subtraction}
\label{sec:skysub_ql}

A Cure routine called \texttt{subtractsky} is called to perform sky subtraction for the science frames. This pipeline provides sky subtraction in two modes. The default and recommended mode uses fibers in the users science frames that are devoid of continuum to model the sky emission. In this mode, each row on the CCD is sampled with a window size defined by the user (typically 100 lines) above and below the line. Within that window, each pixel's flux is divided by its weight from the fiber model and the Jacobian of the transform into wavelength space to account for the changing dispersion along a fiber. Pixels are sorted by their wavelengths and at each approximate wavelength outliers are filtered out. A b-spline is fit along the wavelength direction to these values. Fibers determined to contain continuum based on a signal to noise minimum set by the user are identified and then a second iteration of this process is run with the continuum fibers excluded. The resulting b-splines provide a sky model, for which its value at each pixel's wavelength is multiplied by that pixel's weight, and the Jacobian and then subtracted. \par


The pipeline provides a second option for performing sky subtraction that instead uses fibers in a separate sky frame to build a model of the sky emission. The user has a choice of providing a sky frame or allowing the pipeline to find a parallel frame from another observation at the closest exposure time. Cure's \texttt{subtractsky} routine builds a sky model in the same way as in the previous paragraph using this sky frame.  This second option is not recommended as the sky varies with time and these sky exposures are not taken simultaneously with the users science frames. However, using separate sky frames is sometimes necessary for extremely bright and extended objects where every element of the IFU contains object continuum. \par

\subsection{Fiber Extraction}
\label{sec:fe_ql}

The pipeline offers the user the option to build a fiber extracted file containing the 1D spectra either resampled over a uniform wavelength grid or not. If the user chooses to perform sky subtraction the pipeline chooses the sky subtracted science frames for extraction; otherwise it uses the science frames processed through basic reduction. The pipeline calls Cure's \texttt{fiberextract} routine to perform the extraction. The pipeline specifies this routine to use its profile fitting extraction, in which is uses the fiber model to fit the fluxes of each fiber deblending their signal.  \par

\subsection{Visualization Tools}
\label{sec:vis_ql}

The QLP provides the user with two optional data products to help them visualize their data: data cubes and collapsed cubes. Since LRS2 has a full fill factor, dithers are not necessary, making the construction of data cubes unnecessary for combining data to gain more information. Hence this step is purely a visualization tool, since more information is not obtained by building a data cube.  (Dithers are not combined and the extra resampling actually results in a loss of information.) This tool is provided, however, as it helps the user understand their 3D data, and resampling over a uniform grid allows the spectra to be read with standard fits viewers like ds9 and QFits View. Data cubes are built using Cure's \texttt{mkcube} routine; the pipeline builds the dither file needed by \texttt{mkcube}. The pipeline also defines parameters for the Gaussian kernal interpolation routine specific to the LRS2 PSF.  \par

There is also a collapse cube routine written into pipeline. This routine sums the data cube around a wavelength defined by the user or sums the entire cube if a range is not provided. It returns an image of the collapsed region for visualization. This can be useful to help the use quickly visualize the the spatial distribution of a specific emission line, for example. \par




\bibliographystyle{aasjournal}



\end{document}